\documentclass[twocolumn,twocolappendix]{aastex61.mod}

\newcommand\aastex{AAS\TeX}

\received{March 29, 2018}
\revised{May 10, 2018}
\accepted{May 11, 2018}

\shorttitle{\aastex\ HSTPROMO VI. Improved data reduction and
  internal-kinematic analysis of NGC~362}

\shortauthors{Libralato et al.}

\usepackage{xspace}
\usepackage{amsmath}
\usepackage{rotating}

\newcommand{\magv}{$m_{\rm F606W}$\xspace}
\newcommand{\magi}{$m_{\rm F814W}$\xspace}
\newcommand{\colvi}{$(m_{\rm F606W}-m_{\rm F814W})$\xspace}
\newcommand{\eqmaguv}{m_{\rm F275W}}
\newcommand{\eqmagu}{m_{\rm F336W}}
\newcommand{\eqmagb}{m_{\rm F438W}}
\newcommand{\eqmagv}{m_{\rm F606W}}
\newcommand{\eqmagi}{m_{\rm F814W}}
\newcommand{\eqcolvi}{(m_{\rm F606W}-m_{\rm F814W})}

\defcitealias{2014ApJ...797..115B}{Paper~I}
\defcitealias{2015ApJ...803...29W}{Paper~II}
\defcitealias{2015ApJ...812..149W}{Paper~III}
\defcitealias{2016ApJ...827...12B}{Paper~IV}
\defcitealias{2017ApJ...844..167B}{Paper~V}

\begin{document}

\title{Hubble Space Telescope Proper motion (HSTPROMO) Catalogs of
  Galactic Globular Clusters. VI. Improved data reduction and
  internal-kinematic analysis of NGC~362}

\correspondingauthor{Mattia Libralato}
\email{libra@stsci.edu}

\author[0000-0001-9673-7397]{Mattia Libralato}
\affil{Space Telescope Science Institute 3700 San Martin Drive, Baltimore, MD 21218, USA}
\affil{Dipartimento di Fisica e Astronomia, Universit\`a di Padova, Vicolo dell'Osservatorio 3, Padova, I-35122, Italy}
\affil{INAF-Osservatorio Astronomico di Padova, Vicolo dell'Osservatorio 5, Padova, I-35122, Italy}

\author[0000-0003-3858-637X]{Andrea Bellini}
\affil{Space Telescope Science Institute 3700 San Martin Drive, Baltimore, MD 21218, USA}

\author[0000-0001-7827-7825]{Roeland P. van der Marel}
\affil{Space Telescope Science Institute 3700 San Martin Drive, Baltimore, MD 21218, USA}
\affil{Center for Astrophysical Sciences, Department of Physics \& Astronomy, Johns Hopkins University, Baltimore, MD 21218, USA}

\author[0000-0003-2861-3995]{Jay Anderson}
\affil{Space Telescope Science Institute 3700 San Martin Drive, Baltimore, MD 21218, USA}

\author[0000-0002-1343-134X]{Laura L. Watkins}
\affil{Space Telescope Science Institute 3700 San Martin Drive, Baltimore, MD 21218, USA}

\author[0000-0002-9937-6387]{Giampaolo Piotto}
\affil{Dipartimento di Fisica e Astronomia, Universit\`a di Padova, Vicolo dell'Osservatorio 3, Padova, I-35122, Italy}
\affil{INAF-Osservatorio Astronomico di Padova, Vicolo dell'Osservatorio 5, Padova, I-35122, Italy}

\author[0000-0002-2165-8528]{Francesco R. Ferraro}
\affil{Dipartimento di Fisica e Astronomia, Universit\`a di Bologna, Via Gobetti 93/2, Bologna, I-40129, Italy}
\affil{INAF-Osservatorio Astronomico di Bologna, Via Gobetti 93/3, Bologna, I-40129, Italy}

\author[0000-0003-1149-3659]{Domenico Nardiello}
\affil{Dipartimento di Fisica e Astronomia, Universit\`a di Padova, Vicolo dell'Osservatorio 3, Padova, I-35122, Italy}
\affil{INAF-Osservatorio Astronomico di Padova, Vicolo dell'Osservatorio 5, Padova, I-35122, Italy}

\author[0000-0003-2742-6872]{Enrico Vesperini}
\affil{Department of Astronomy, Indiana University, Bloomington, IN 47405, USA}

\begin{abstract}

We present an improved data-reduction technique to obtain
high-precision proper motions (PMs) of globular clusters using
\textit{Hubble Space Telescope} data. The new reduction is superior to
the one presented in the first paper of this series for the faintest
sources in very crowded fields.  We choose the globular cluster
NGC~362 as a benchmark to test our new procedures. We measure PMs of
117\,450 sources in the field, showing that we are able to obtain a PM
precision better than 10 $\mu$as yr$^{-1}$ for bright stars. We make
use of this new PM catalog of NGC 362 to study the cluster's internal
kinematics. We investigate the velocity-dispersion profiles of the
multiple stellar populations hosted by NGC~362 and find new pieces of
information on the kinematics of first- and second-generation stars.
We analyze the level of energy equipartition of the cluster and find
direct evidence for its post-core-collapsed state from kinematic
arguments alone.  We refine the dynamical mass of the blue stragglers
and study possible kinematic differences between blue stragglers
formed by collisions and mass transfer. We also measure no significant
cluster rotation in the plane of the sky. Finally, we measure the
absolute PM of NGC~362 and of the background stars belonging to the
Small Magellanic Cloud, finding a good agreement with previous
estimates in the literature. We make the PM catalog publicly
available.

\end{abstract}


%%%%%%%%
\section{Introduction}
%%%%%%%%

High-precision proper motions (PMs) have proven to be the most
effective tool to analyze the internal kinematics and dynamics of
globular clusters (GCs). Although new instruments and missions have
been recently developed in this context, the \textit{Hubble Space
  Telescope} (\textit{HST}) is still the reference astrometric tool
for such investigations.

\citet[Paper~I]{2014ApJ...797..115B} computed the PM of stars in 22
GCs using archival \textit{HST} data sets. These PM catalogs, which
represent the state-of-the-art for astrometry, make it possible to
investigate a broad range of kinematic studies. For example,
\citet[Paper~II]{2015ApJ...803...29W} and
\citet[Paper~III]{2015ApJ...812..149W} used these PMs to compute the
velocity dispersions for cluster stars at different radial distances,
two-dimensional velocity-dispersion spatial maps, dynamical distances
and mass-to-light ratios for these GCs.  Furthermore, by means of the
same PM catalogs, \citet[Paper~IV]{2016ApJ...827...12B} computed the
dynamical mass of the blue stragglers (BSs).

These catalogs collect the astrometric and photometric data for all
objects in the observed fields of these GCs. As such,
foreground/background sources are measured with the same astrometric
precision as GC stars and can be adopted to infer the cluster's
rotation in the plane of the sky of the GCs, as was done by
\citet[Paper~V]{2017ApJ...844..167B} for NGC~104 (47\,Tuc).

In this paper of the series, we derive a new PM catalog for the GC
NGC~362 by using a revised data-reduction strategy, which was
specifically designed to improve PM precision and completeness in
very-crowded fields. Since the procedures are somewhat complicated, we
describe them in detail in the Appendices.

In Appendix~\ref{red} we describe the photometric procedures.  A
one-pass photometric routine is run to measure the bright, easy to
measure stars, and these measurements are used to determine the
photometric and astrometric transformations between each exposure and
the reference frame.  We then introduce a second stage of reduction
that (i) simultaneously employs all images at once, and (ii) applies
neighbor subtraction. The former feature works best in enhancing the
contribution of faint sources otherwise lost in the noise of the
images. The latter is particularly effective near the center of
GCs. The core of NGC~362 is extremely crowded, thus being a perfect
benchmark to highlight the improvements made possible by the new
data-reduction strategy.

In Appendix \ref{rpm} describe how we compute the relative proper
motions for all the stars.

NGC~362 is a core-collapsed GC \citep[e.g.,][]{1995AJ....109..218T}
that is known to host multiple stellar populations \citep[hereafter
  mPOPs, see,
  e.g.,][]{2012ApJ...760...39P,2013A&A...557A.138C,2016ApJ...832...99L,2015AJ....149...91P,2017MNRAS.464.3636M},
as well as two sequences of BSs \citep{2013ApJ...778..135D}. We
investigate the internal kinematics of this GC, in particular of its
mPOPs, the level of energy equipartition, and, for the first time, its
post-core-collapsed state thanks solely to internal-kinematic
arguments (Sect.~\ref{analysis}).

Finally, NGC~362 is located in front of the Small Magellanic Cloud
(SMC). As such, we can estimate the rotation of NGC~362 in the plane
of the sky using the same method described in
\citetalias{2017ApJ...844..167B}.

%%%%%%%%
\section{Data sets and reduction}\label{obs}
%%%%%%%%

\begin{figure}
  \centering
  \includegraphics[width=\columnwidth]{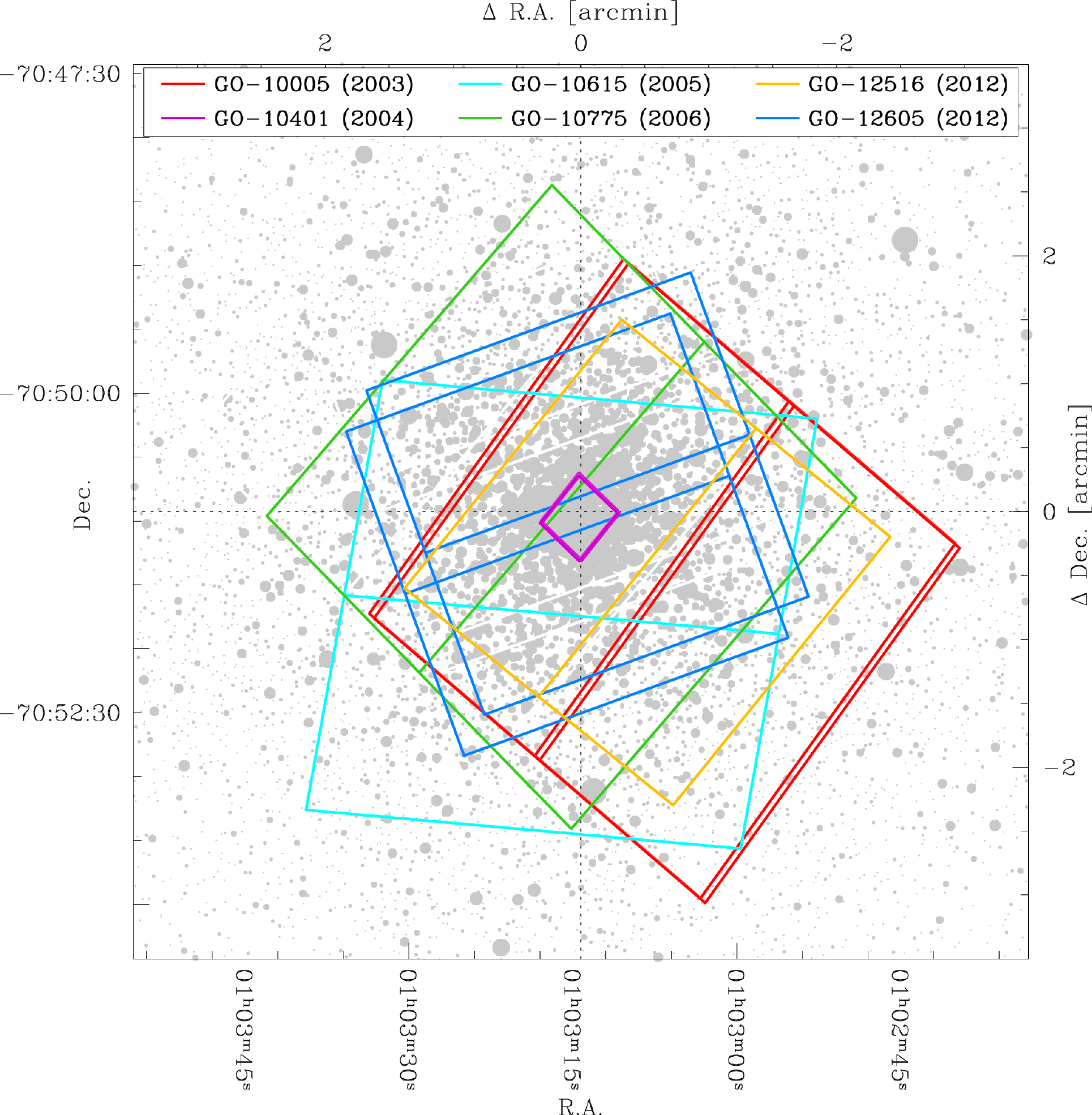}
  \caption{Outline of the FoV covered by the observations used for the
    analysis of NGC~362. Each footprint is color-coded according to
    the GO proposal number. The stellar map is obtained using the
    $G$-band photometry of the Gaia Data Release 1
    \citep[DR1,][]{2016A&A...595A...1G,2016A&A...595A...2G}. Top and
    right axes are in arcmin with respect to the cluster center given
    by \citet{2010AJ....140.1830G}.}
  \label{fig:fov}
\end{figure}

We made use of all suitable \textit{HST} images covering the central
field of the cluster\footnote{$(\alpha,\delta)_{\rm J2000}$ $=$
  ($01^{\rm h}03^{\rm m}14^{\rm
    s}\!\!.26$,$-70^\circ50^\prime55^{\prime\prime}\!\!.6$),
  \citet{2010AJ....140.1830G}.}. We focused on High-Resolution Channel
(HRC) and the Wide-Field Channel (WFC) exposures taken with the
Advanced Camera for Survey (ACS), and with the Ultraviolet-VISible
(UVIS) channel of the Wide-Field Camera 3 (WFC3). Observations of the
Wide-Field Planetary Camera 2 (WFPC2) were not used because of the
much lower astrometric precision with respect to that of ACS and WFC3
(UVIS) imagers. An overview of the field of view (FoV) covered is
shown in Fig.~\ref{fig:fov}. Table~\ref{tab:log} lists all the
observations analyzed in this paper\footnote{DOI reference:
  \protect\dataset[10.17909/T9CH53]{http://dx.doi.org/10.17909/T9CH53}}.

\begin{table*}
  \caption{List of observations of NGC~362 used in this paper.}
  \centering
  \label{tab:log}
  \begin{tabular}{cccccc}
    \hline
    \hline
    GO & PI & Instrument/Camera & Filter & $N$ $\times$ Exp. Time & Epoch \\
    \hline
    10005 & Lewin & ACS/WFC & F435W & $4 \times 340$ s & 2003 December \\
    & & & F625W & $2 \times 110$ s , $2 \times 120$ s & \\
    & & & F658N & $2 \times 440$ s , $2 \times 500$ s & \\
    10401 & Chandar & ACS/HRC & F435W & $16 \times 85$ s & 2004 December \\
    10615 & Anderson & ACS/WFC & F435W & $30 \times 340$ s , $5 \times 70$ s & 2005 September \\
    10775 & Sarajedini & ACS/WFC & F606W & $4 \times 150$ s , $1 \times 10$ s & 2006 June \\
    & & & F814W & $4 \times 170$ s , $1 \times 10$ s & \\
    12516 & Ferraro & WFC3/UVIS & F390W & $14 \times 348$ s & 2012 April \\
    & & & F555W & $1 \times 160$ s , $1 \times 200$ s \\
    & & &       & $1 \times 144$ s , $1 \times 145$ s \\
    & & &       & $6 \times 150$ s \\
    & & & F814W & $12 \times 348$ s , $3 \times 390$ s & \\
    12605 & Piotto & WFC3/UVIS & F275W & $6 \times 519$ s & 2012 September \\
    & & & F336W & $4 \times 350$ s & \\
    & & & F438W & $4 \times 54$ s & \\
    \hline
  \end{tabular}
\end{table*}

The data reduction was performed closely following the prescriptions
extensively described in
\citet{2017ApJ...842....6B,2018ApJ...853...86B} and it is the end
result of first- and second-pass photometric reductions. PMs were
computed as described in \citetalias{2014ApJ...797..115B} and
\citet{2018ApJ...853...86B}, an advanced evolution of the so-called
central overlap method \citep{1971PMcCO..16..267E}. An overview of the
procedures adopted and a comparison with the original PM catalog of
\citetalias{2014ApJ...797..115B} are presented in Appendices~\ref{red}
and \ref{rpm}.

\begin{figure*}
  \centering
  \includegraphics[width=\textwidth]{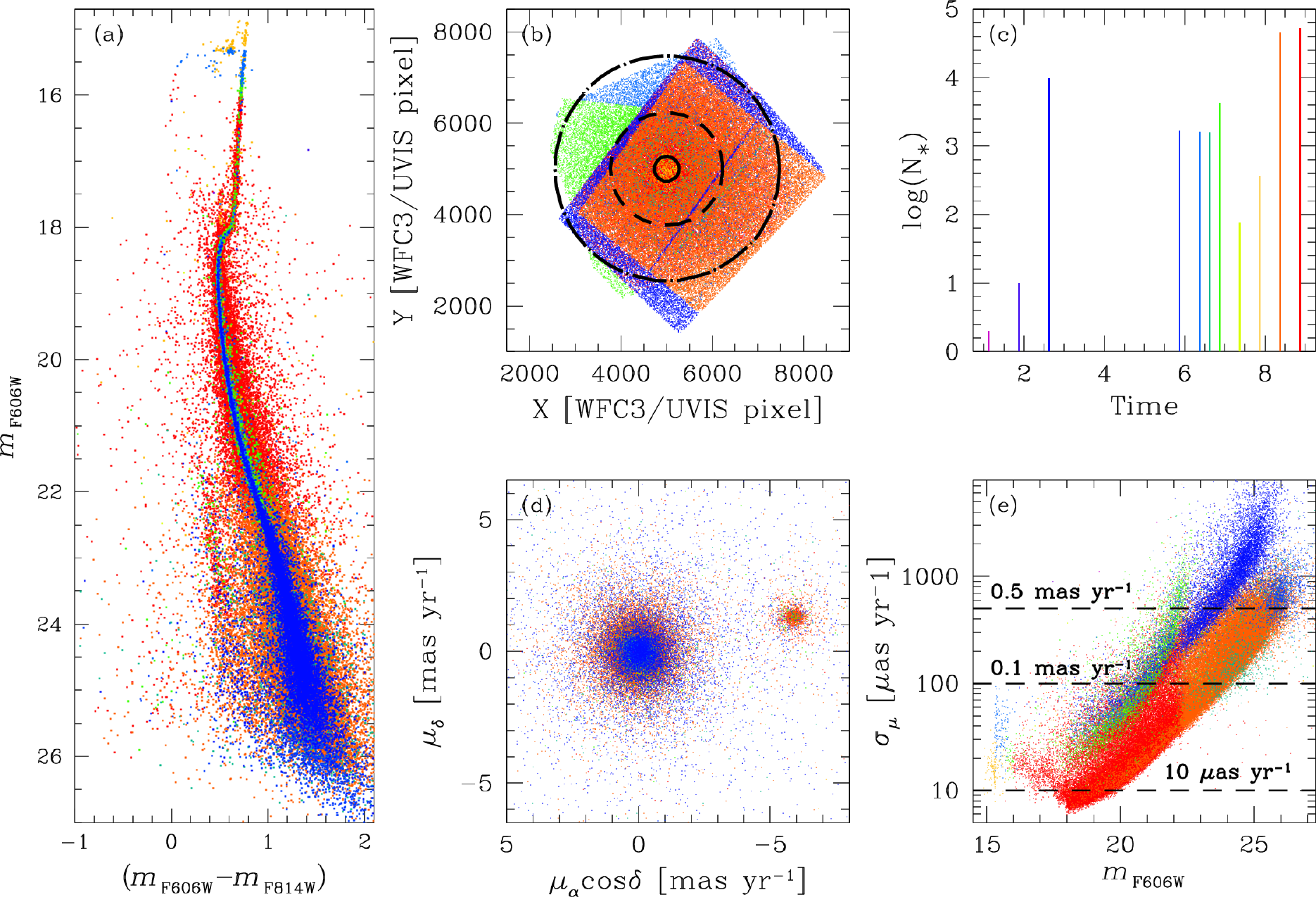}
  \caption{Overview of the PM catalog constructed in this paper. The
    \magv versus \colvi CMD of NGC~362 is presented in panel (a). In
    panel (b) we show the field covered by the analyzed data sets in
    units of WFC3/UVIS pixels. The cluster is centered at (5000,5000)
    WFC3/UVIS pixels. The solid circle has a radius equal to the
    $r_{\rm c}$ of NGC~362 \citep[0.18 arcmin,][2010
      edition]{1996AJ....112.1487H}, while the short-dashed and the
    long-dashed circles have a radius equal to $r_{\rm h}$ and
    $2r_{\rm h}$, respectively. The histogram in panel (c) shows the
    logarithm of the number of sources as a function of the temporal
    baseline. The points in all panels of this Figure are color-coded
    according to the temporal baseline as shown in panel (c). The
    relative VPD in equatorial coordinates is shown in panel
    (d). Panel (e) shows the 1-D PM error err$_\mu$ in $\mu$as
    yr$^{-1}$ as a function of the \magv magnitude. The three dashed
    lines are at 10, 100 and 500 $\mu$as yr$^{-1}$. In panels (d) and
    (e) we show the original, uncorrected PMs (see the text for
    details).}
  \label{fig:overview}
\end{figure*}

The final PM catalog contains 117450 sources. An overview of the PM
catalog is shown in Fig.~\ref{fig:overview}. In panel (a) we present
the color-magnitude diagram (CMD) of NGC~362. We were able to measure
PMs from the red-giant branch (RGB) down to $\sim 8$ magnitudes below
the main-sequence (MS) turn-off. In panel (b) we show the FoV covered
by our data sets. In all panels, sources are color-coded according to
the temporal baseline used to compute their PM (see histogram in panel
c). The stellar PMs were computed with a temporal baseline between
about 1 and 9 yr. Stars within the half-light radius \citep[$r_{\rm h}
  = 0.82$ arcmin,][2010 edition]{1996AJ....112.1487H} have PMs
computed over a $\sim 9$-yr baseline, while outside of $r_{\rm h}$ the
available temporal coverage is shorter. Two groups of stars are
clearly distinguishable in the vector-point diagram (VPD) of panel
(d):\ NGC~362 stars are centered at the origin of the VPD, by
construction;\ SMC stars populate the lesser clump around $(-5.9,1.3)$
mas\,yr$^{-1}$. In panel (e) we show the 1-D PM error as a function of
\magv magnitude. The PM of bright, well-measured stars in the deep
exposures is precise to better than 10 $\mu$as\,yr$^{-1}$\footnote{The
  effect of the annual parallax can be as large as $\sim 26$ $\mu$as
  yr$^{-1}$ for stars that are members of NGC~362, a factor two
  greater than the PM error for well-measured, bright stars. However,
  we computed the PMs using cluster stars as reference. The parallax
  effect has no impact on NGC~362 stars but it is instead transferred,
  with opposite sign, to foreground/background objects. Therefore, the
  analyses of the internal kinematics of NGC~362 in our paper should
  not be affected by annual-parallax effects.}. Faint stars at about
$\eqmagv \sim 26$ are measured with a precision of 0.5 mas
yr$^{-1}$. As a reference, the expected end-of-mission precision of
Gaia at $\eqmagv \sim 21.5$ (the Gaia faint limit) is $\sim 1$
mas\,yr$^{-1}$ \citep[see][]{2017MNRAS.467..412P}; our PM error at
this magnitude is about 0.042 mas yr$^{-1}$, with a few stars having
errors around 0.020 mas yr$^{-1}$. At the faint end PMs are also
measured reasonably well (i.e., err$_\mu < 0.5$ mas yr$^{-1}$ at
$\eqmagv \sim 25$), even when a small number of exposures is
available. This is thanks to the neighbor-subtraction characteristics
of \texttt{KS2} (see Appendix~\ref{red}).

We also made a comparison with the PMs recently available from the
Gaia DR2 \citep{2016A&A...595A...1G,2018arXiv180409365G}. We computed
the median 1-D PM error err$_\mu$ as a function of magnitude using
only the stars in common between the two catalogs and found that the
precision of our PMs is between a factor 4 (bright stars with $G \sim
15.4$) and 85 (faint stars with $G>19$) times better than that of the
Gaia DR2. Furthermore, we found that the $\sim$75\% of the common
stars are beyond 1 arcmin from the center, as might be expected
because of the level of crowding in the field.

Two sequences of errors are visible in panel (e) of
Fig.~\ref{fig:overview}, hereafter the ``best'' and the ``worst''
sequences. The worst sequence seems to also be split in two branches,
one brighter than $\eqmagv \sim 23$ and one fainter. The faint tail of
the worst sequence is related to stars which PM is measured with a
temporal baseline lower than 3 yr.  Furthermore, the majority of stars
belonging to the worst-measured sequences have PMs computed with less
than 30 and 50 images, respectively.

Finally, we investigated the presence of positional-, magnitude- and
color-dependent systematic errors in our PMs and found no clear
trends, giving us assurance that our PMs are not affected by
large-scale systematic effects.

%%%%%%%%
\section{Analysis}\label{analysis}
%%%%%%%%

%%%%%%%%%%%
\subsection{PM selections and validation}\label{kincheck}
%%%%%%%%%%%

The analysis of the internal kinematics of GCs requires a careful
selection of the best-measured stars. In general, we adopted the
following constraints on stars to be considered: (i) the reduced
$\chi^2$ of the PM fit must be lower than 1.5, (ii) PM-fitting
rejection rate must be lower than 15\%, and (iii) the quality of fit
(\textsf{QFIT}) value (a measure of how well the star is fit by a
single-star PSF) must be greater than the 50$^{\rm th}$-percentile
\textsf{QFIT} value at the star's magnitude level (see
Appendices~\ref{red} and \ref{rpm} for a more complete description of
these quantities).

In addition, we kept cluster stars with a PM error lower than half the
local velocity dispersion $\sigma_\mu$ of the closest (in distance and
magnitude) 75 cluster stars (see Sect.~8.3 of
\citetalias{2014ApJ...797..115B}). We considered as cluster members
stars with a relative PM lower than 1.2 mas yr$^{-1}$ (about 5 times
the velocity dispersion of faint members of NGC~362).

To analyze how different \textsf{QFIT} cuts could change the inferred
velocity-dispersion profile of NGC~362, we computed the value of
$\sigma_\mu$ in different radial and magnitude bins for \textsf{QFIT}
thresholds from the 5$^{\rm th}$ to the 95$^{\rm th}$ percentile, with
steps of 5\%. We found that the values of $\sigma_\mu$ in our tests
are all consistent with each other, mainly because most of the
outliers are already removed with the reduced $\chi^2$ and the
rejection-rate selections. As such, we arbitrarily chose the 50$^{\rm
  th}$ percentile.

\begin{figure*}
  \centering
  \includegraphics[width=0.85\textwidth]{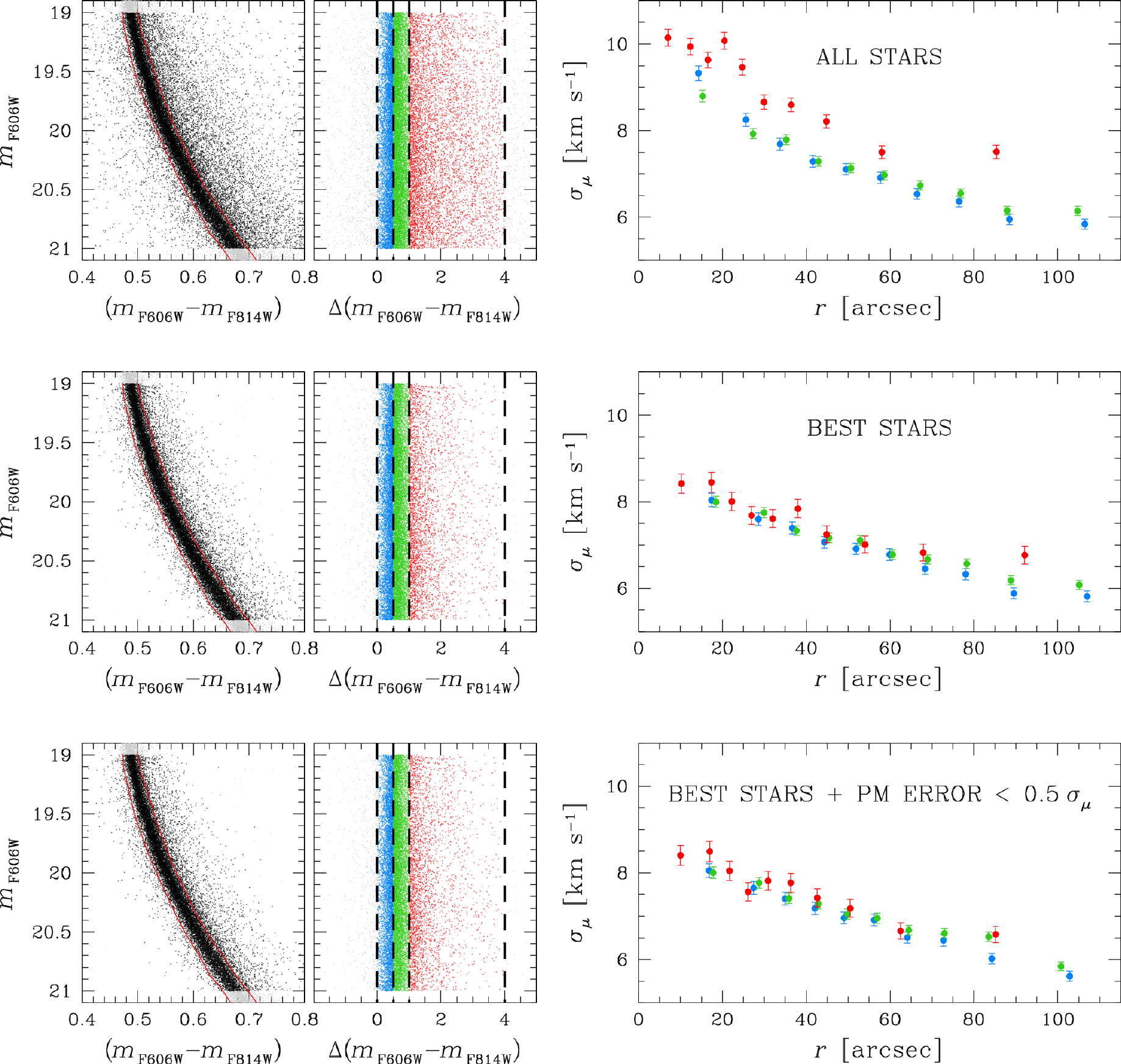}
  \caption{On the left we present the $m_{\rm F606W}$ versus \colvi
    CMD of MS stars of NGC~362 with $19 < \eqmagv < 21$. The red
    fiducial lines are used to rectify the CMD (central panels). The
    dashed, vertical black lines define three samples of stars that
    are used to compute velocity-dispersion radial profiles. In the
    right-hand panels, we show the combined velocity-dispersion
    profiles as a function of distance from the cluster center for
    these three groups. From top to bottom, we present the analysis
    obtained by considering all stars, well-measured stars defined as
    described in the text, and a sub-sample of well-measured stars
    with PM errors smaller than half of the average velocity
    dispersion, respectively. See the text for details}
  \label{fig:testo2}
\end{figure*}

\begin{figure*}
  \centering
  \includegraphics[width=\textwidth]{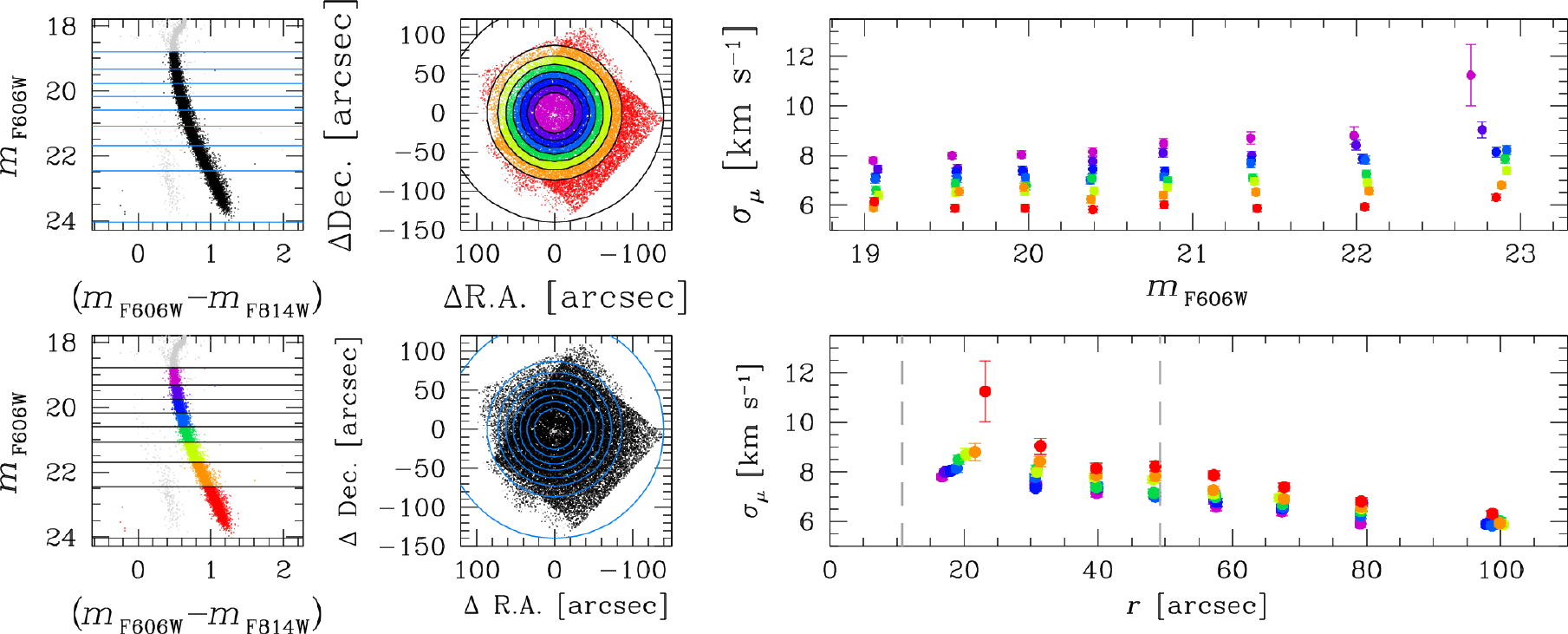}
  \caption{Velocity-dispersion profiles of MS stars of
    NGC~362. (Top-left): \magv versus \colvi CMD. Black points
    represent NGC~362 members that survived the quality selections
    described in the text. Gray dots are non-MS objects. We considered
    only stars from the MS turn-off ($\eqmagv \sim 18.8$) down to five
    magnitudes below (about the faintest magnitude of the stars in the
    sample, $\eqmagv \sim 24$). The azure lines define the
    magnitude-bin limits of 8 equally-populated intervals adopted in
    the analysis. (Top-middle): FoV of selected stars. The black
    circles define 8 equally-populated radial bins. Stars are
    color-coded according to the radial interval they belong
    to. (Top-right): $\sigma_\mu$ as a function of \magv. Points (with
    error bars) are color-coded as in top-middle panel. (Bottom): as
    on the top panels, but for $\sigma_\mu$ in different \magv bins as
    a function of radial distance.}
  \label{fig:any2rmpaper}
\end{figure*}

In Fig.~\ref{fig:testo2} we show the velocity-dispersion radial
profile inferred from the PMs measured in our catalog with different
selection criteria. We considered only MS members of NGC~362 with $19
< \eqmagv < 21$. We drew by hand two fiducial lines enclosing the bulk
of MS stars and made use of these lines to rectify the MS. We then
split the MS in three groups: ``blue'' stars with $0 \leq \Delta
\eqcolvi < 0.5$ (azure points), ``red'' sources with $0.5 \leq \Delta
\eqcolvi < 1.0$ (green points), and ``very-red'' objects $1.0 \leq
\Delta \eqcolvi < 4.0$ (red points). Finally, for each group we
computed the velocity dispersion $\sigma_\mu$ in 10 equally-populated
radial bins\footnote{Binning the data will always introduce biases,
  some of which can be quite subtle. We applied different criteria for
  the bin sizes and found consistent results within the errors.}. The
velocity dispersion is computed as described in
\citet{2010ApJ...710.1063V}, i.e., by correcting the observed scatter
of the PMs for the uncertainties of the individual PMs.

Without any selection (top panels), very-red objects appear
kinematically hotter than blue and red stars. Most of these sources
are blends or poorly-measured stars due to crowding, with some
contamination due to binaries \citep[the fraction of MS binaries in
  NGC~362 is less than 5\%,][]{2012A&A...540A..16M}. As such, a
systematic component is present in the PM of these objects, resulting
in a faster \citep[by 0.025 mas yr$^{-1}$ or $\sim 1$ km s$^{-1}$
  assuming a distance of 8.6 kpc,][2010 edition]{1996AJ....112.1487H}
motion with respect to blue and red sources.

In the middle panels of Fig.~\ref{fig:testo2}, we show the
velocity-dispersion profile obtained by considering only the best
stars defined using the PM $\chi^2$, the rejection rate and the
F606W/F814W \textsf{QFIT} selections. The agreement between the three
groups is now improved, with only a marginal departure at large radii.

Finally, in the bottom panels we present the velocity dispersions of
the three groups after we further add a selection on the PM error
based on the local value of $\sigma_\mu$. By removing additional
outliers, the three $\sigma_\mu$ trends agree even in the outermost
part of the field.

At a given radial distance, faint, less-massive MS stars move faster
(are kinematically hotter) than bright, more-massive MS stars in
accordance with what we would expect from energy equipartition.  In
addition, stars of similar mass closer to the center of the cluster
are expected to move faster than those far from the center because of
hydrostatic equilibrium. To examine these two effects, we selected a
sample of MS stars from about the MS turn-off ($\eqmagv \sim 18.8$) to
five magnitudes below. We then divided the sample in 8
equally-populated bins in magnitude and radial distance, and computed
the combined velocity dispersion in each bin.

In Fig.~\ref{fig:any2rmpaper} we summarize the results. In the top
panel, the behavior induced by energy equipartition is clear: the
brighter the star, the heavier we would expect it to be, and hence the
slower its motion. In the bottom panel, stars far from the center of
NGC~362 are kinematically colder than those near the center at a given
magnitude (mass), as expected from hydrostatic equilibrium.

%%%%%%%%%%%
\subsection{mPOP kinematics}\label{mpop}
%%%%%%%%%%%

NGC~362 is known to host mPOPs along its RGB and sub-giant branch
\citep[SGB; see
  e.g.,][]{2012ApJ...760...39P,2013A&A...557A.138C,2016ApJ...832...99L}. In
the following analyses, we considered only the best stars from the
photometric (using the prescriptions given in Appendix~\ref{corr}) and
astrometric (see Sect.~\ref{kincheck}) points of view. The photometric
selections were applied to all F275W, F336W, F438W, F606W and F814W
filters, and are used to identify the different mPOPs in the CMD.

%%%%%%%%%%%%%%
\subsubsection{Red-giant branch}\label{rgb}
%%%%%%%%%%%%%%

%%%%%%%%%%
\paragraph{mPOP tagging}
%%%%%%%%%%

Spectroscopic and photometric studies of the RGB stars of NGC~362 have
revealed peculiar features, i.e., an anti-correlation in the Na-O
plane and different Ba abundances, with Ba-rich stars populating a
distinct RGB in the Str\"omgren $v$ versus $(v-y)$ CMDs
\citep[see][and references therein]{2013A&A...557A.138C}. Furthermore,
\citet{2016ApJ...832...99L} found that CN-weak and CN-strong stars
populated distinct sequences along the RGB in CMDs based on
narrow-band photometry.

More recently, as part of the project ``\textit{Hubble Space
  Telescope} UV Legacy Survey of Galactic GCs''
\citep{2015AJ....149...91P}, \citet{2017MNRAS.464.3636M} analyzed the
RGB of 57 Galactic GCs, classifying NGC~362 as a type-II cluster. The
RGBs of type-II GCs present (pseudo) two-color diagrams (``chromosome
maps'') with split first- (1G) and second-generation (2G)
sequences. Some of the mPOPs of these type-II GCs are also enriched in
the total C$+$N$+$O abundance, the iron content and the abundance of
$s$-process elements (such as Ba).

\begin{figure*}
  \centering
  \includegraphics[width=0.8\textwidth]{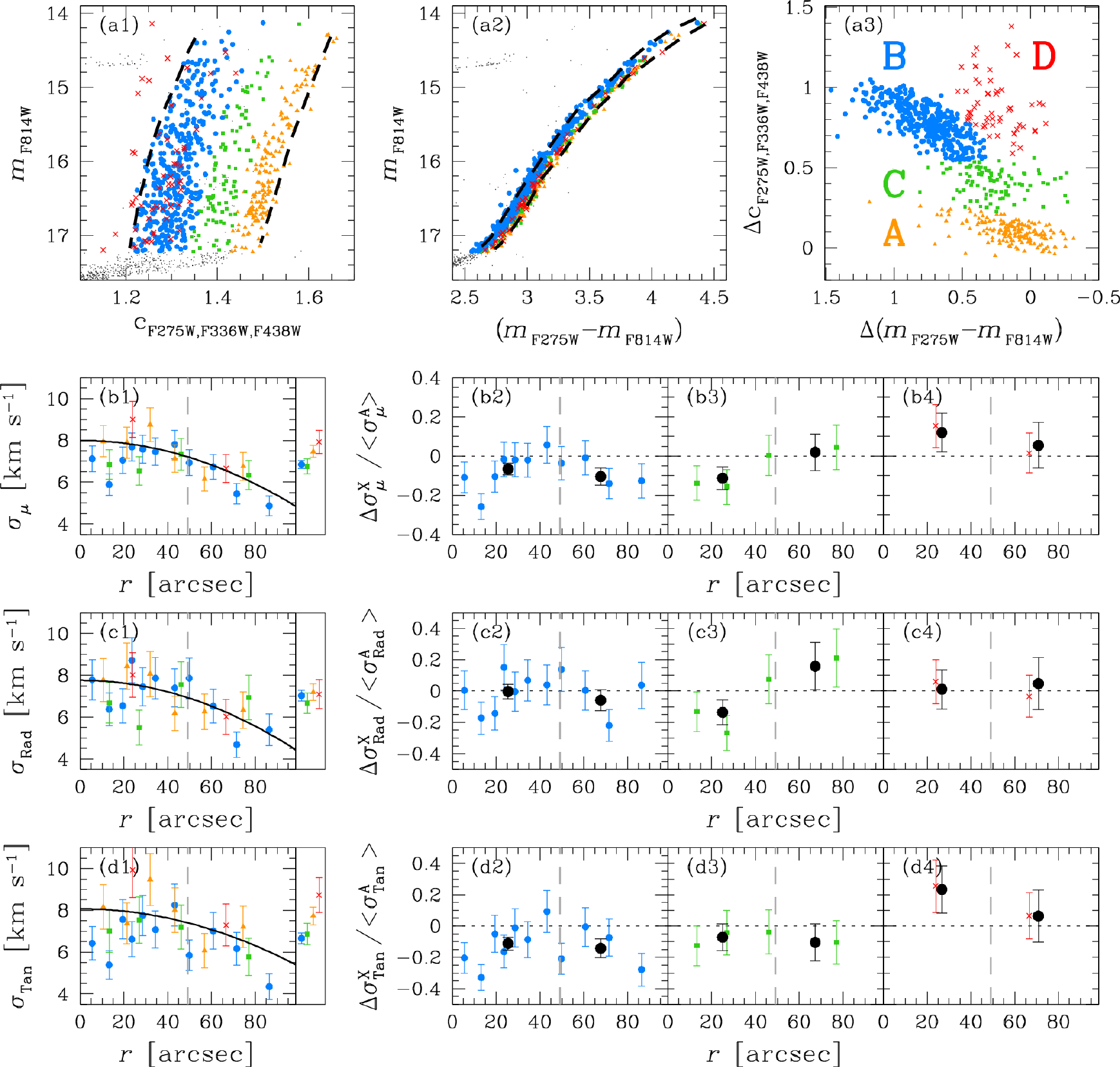}
  \caption{(a1): \magi versus c$_{\rm 275W,F336W,F438W}$ pseudo CMD of
    RGB stars. (a2): \magi versus $(\eqmaguv-\eqmagi)$ CMD of the same
    stars. The two black, dashed lines are used to construct the
    chromosome map. (a3) chromosome map of RGB stars. In these panels,
    points are color-coded according to the 4 mPOPs A (yellow
    triangles), B (azure dots), C (green squares), and D (red
    crosses). (b1): $\sigma_\mu$ as a function of radial
    distance. Points are color-coded as in CMDs of panels (a). The
    solid, black line represents a 3$^{\rm rd}$-order polynomial fit
    to the population A data (yellow crosses), chosen as a
    reference. The gray, dashed vertical line marks the half-light
    radius ($r_{\rm h}$). Points in the inset to the right are the
    average $\sigma_\mu$ over the entire FoV. (b2-b3-b4): normalized
    difference between $\sigma_\mu$ of the 2G populations with respect
    to that of the reference 1G population. The gray, horizontal
    dashed lines are set at 0. The two black dots in each panel are
    the average $\sigma_\mu$ for stars with $r<r_{\rm h}$ and $r_{\rm
      h} \leq r < 2r_{\rm h}$ (dashed, gray vertical
    line). (c1-c2-c3-c4): As in panels (b) but for the radial velocity
    dispersion $\sigma_{\rm Rad}$. (d1-d2-d3-d4): As in panels (b) and
    (c) but for the tangential velocity dispersion $\sigma_{\rm
      Tan}$.}
  \label{fig:rgbpaper1}

  \centering
  \includegraphics[width=0.8\textwidth]{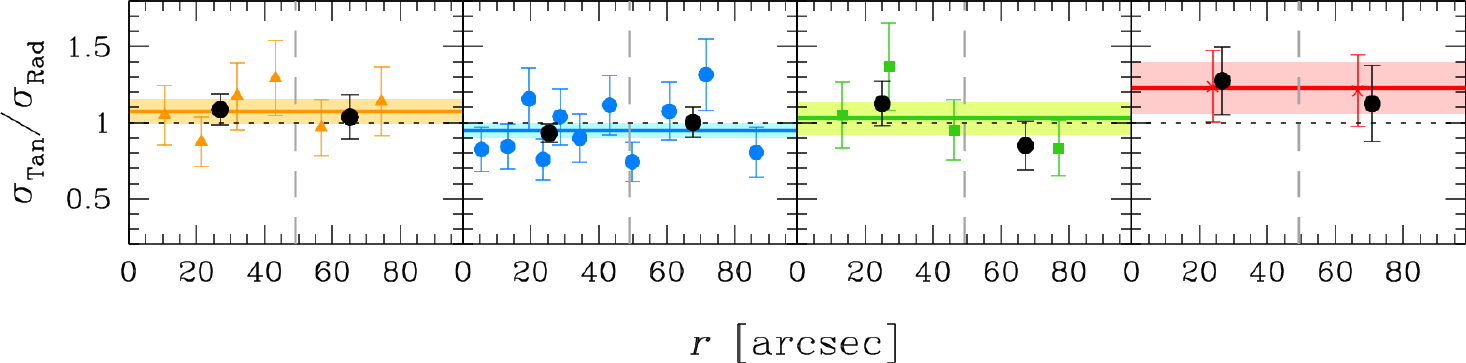}
  \caption{Tangential-to-radial anisotropy profile of the four
    populations on the RGBs of NGC~362 (defined as in
    Fig.~\ref{fig:rgbpaper1}). In each panel, points have the same
    colors and meaning as in Fig.~\ref{fig:rgbpaper1}. The horizontal
    lines represent the average over the entire FoV, the shaded
    regions correspond to $\pm 1\sigma$ errors. The gray, dashed
    vertical line is set at $r_{\rm h}$.}
  \label{fig:rgbpaper2}
\end{figure*}

To separate the different mPOPs in NGC~362, we followed the approach
of \citet{2017MNRAS.464.3636M}. First, in the \magi versus
$\rm{c}_{\rm F275W,F336W,F438W} =
(\eqmaguv-\eqmagu)-(\eqmagu-\eqmagb)$ CMD (panel a1 of
Fig.~\ref{fig:rgbpaper1}), we drew by hand two fiducial lines at the
blue and red edges of the RGBs. These lines were then used to rectify
the RGBs and compute the pseudo-color $\Delta {\rm c}_{\rm
  F275W,F336W,F438W} = ({\rm c}_{\rm
  F275W,F336W,F438W}-{\rm{fiducial_{red}}})/({\rm{fiducial_{blue}}}-{\rm{fiducial_{red}}})$. We
performed the same procedure in the \magi versus $(\eqmaguv-\eqmagi)$
CMD (panel a2) to compute the pseudo-color $\Delta
(\eqmaguv-\eqmagi)$. The resulting chromosome map in the $\Delta {\rm
  c}_{\rm F275W,F336W,F438W}$ versus $\Delta (\eqmaguv-\eqmagi)$ plane
is shown in panel (a3) of Fig.~\ref{fig:rgbpaper1}.

As described in \citet{2017MNRAS.464.3636M}, 1G stars are expected to
be located at $\Delta {\rm c}_{\rm F275W,F336W,F438W} \sim 0$ (yellow
triangles), while the remaining stars belong to the 2G. The red
crosses represent the so-called red-RGB stars, stars with a higher
C$+$N$+$O, Fe and $s$-process-element abundances than other stars of
the same generation. These red-RGB stars are about the 9\% of the
total RGB stars in our sample, in agreement with the analysis of
\citet{2017MNRAS.464.3636M}. The remaining stars seem to be split in
two groups in the chromosome map (azure dots and green squares).

The 1G/2G nature of these four groups was confirmed by
cross-correlating our first-pass photometric catalog (which includes
the saturated stars, even though we cannot measure their PMs) with the
spectroscopic catalog of of \citet{2013A&A...557A.138C}. We found that
the azure- and red-RGB stars are Na rich and O poor, as expected by 2G
stars, while the green-RGB stars have a chemistry of either 1G or 2G
stars. Finally, the yellow RGB stars are located in the Na-poor/O-rich
region of the plot, as expected by 1G stars.

Hereafter, we refer to the 1G yellow-, 2G azure-, 2G green-, and 2G
red-RGB populations as populations A, B, C, and D, respectively.

%%%%%%%%%%
\paragraph{Internal kinematics}
%%%%%%%%%%

In Fig.~\ref{fig:rgbpaper1} we present the analysis of the internal
kinematics of the mPOPs identified on the RGB of NGC~362. As in
Sect.~\ref{kincheck}, we corrected the observed scatter of the PMs for
the PM errors. In panel (b1), we show the combined velocity dispersion
$\sigma_\mu$ as a function of the radial distance. We inferred the
value of $\sigma_\mu$ in bins of 29, 33, 24, and 27 star each for
populations A, B, C, and D, respectively. We then chose as a reference
population A and fitted the data with a 3$^{\rm rd}$-order polynomial
(black line). In the inset of panel (b1), we depicted the global
$\sigma_\mu$ of all stars in our FoV as reference. In panels (b2),
(b3), and (b4) we show the normalized difference between $\sigma_\mu$
of 2G and 1G stars. In each panel, black points were computed by using
stars within $r_{\rm h}$ or between $r_{\rm h} \leq r < 2r_{\rm
  h}$. In panels (c) and (d) of Fig.~\ref{fig:rgbpaper1}, we show the
radial ($\sigma_{\rm Rad}$) and tangential ($\sigma_{\rm Tan}$)
velocity dispersion as a function of the radial distance and the
normalized difference between 2G and 1G mPOPs as in panels (b).

We find that 1G and 2G stars exhibit the same kinematics. There is
only marginal evidence of population B having a lower $\sigma_\mu$
than population A at the $\sim 2.2\sigma$ level because of a lower
$\sigma_{\rm Tan}$.

We also computed the radial anisotropy profile for these RGB stars. In
Fig.~\ref{fig:rgbpaper2} we show the ratio $\sigma_{\rm
  Tan}/\sigma_{\rm Rad}$ as a function of distance from the center of
NGC~362 for each mPOP on the RGB. In each panel, the horizontal line
represents the average trend of the population computed using all
stars in the field. The shaded area indicates the $\pm 1\sigma$ error
bars. On average, both 1G and 2G stars are consistent with an
isotropic system.

The smaller tangential velocity dispersion of the 2G population B is
similar, although less evident, to the observational findings and
theoretical simulations of 47\,Tuc \citep{2013ApJ...771L..15R},
NGC~2808 \citep{2015ApJ...810L..13B}, and $\omega$\,Cen
\citep{2018ApJ...853...86B}. However, the difference between
$\sigma_{\rm Rad}$ and $\sigma_{\rm Tan}$ in the RGB population B of
NGC~362 is not large enough to create a significant radial anisotropy
as in the case of NGC~2808 and $\omega$\,Cen.

%%%%%%%%%%%%%%
\subsubsection{Sub-giant branch}\label{sgb}
%%%%%%%%%%%%%%

Like the RGB, the SGB of NGC~362 is also split in at least 2 main
groups \citep{2012ApJ...760...39P}. The less-populated group is
clearly separated from the remaining SGB stars in the $\eqmagi$ versus
$(\eqmagu-\eqmagi)$ CMD. As suggested by \citet{2017MNRAS.464.3636M},
this SGB is connected to the red-RGB population D.

We initially identified the stars belonging to the less-populated SGB
(hereafter, the red SGB) in the $\eqmagi$ versus $(\eqmagu-\eqmagi)$
CMD. These stars are about 9\% of the SGB stars in our sample. Then,
we analyzed in detail the more-populated SGB by constructing a
chromosome map as we did for the RGB stars. Unlike the RGB
investigation, we made use of a UV-filter-only map.

\begin{figure*}
  \centering
  \includegraphics[width=0.8\textwidth]{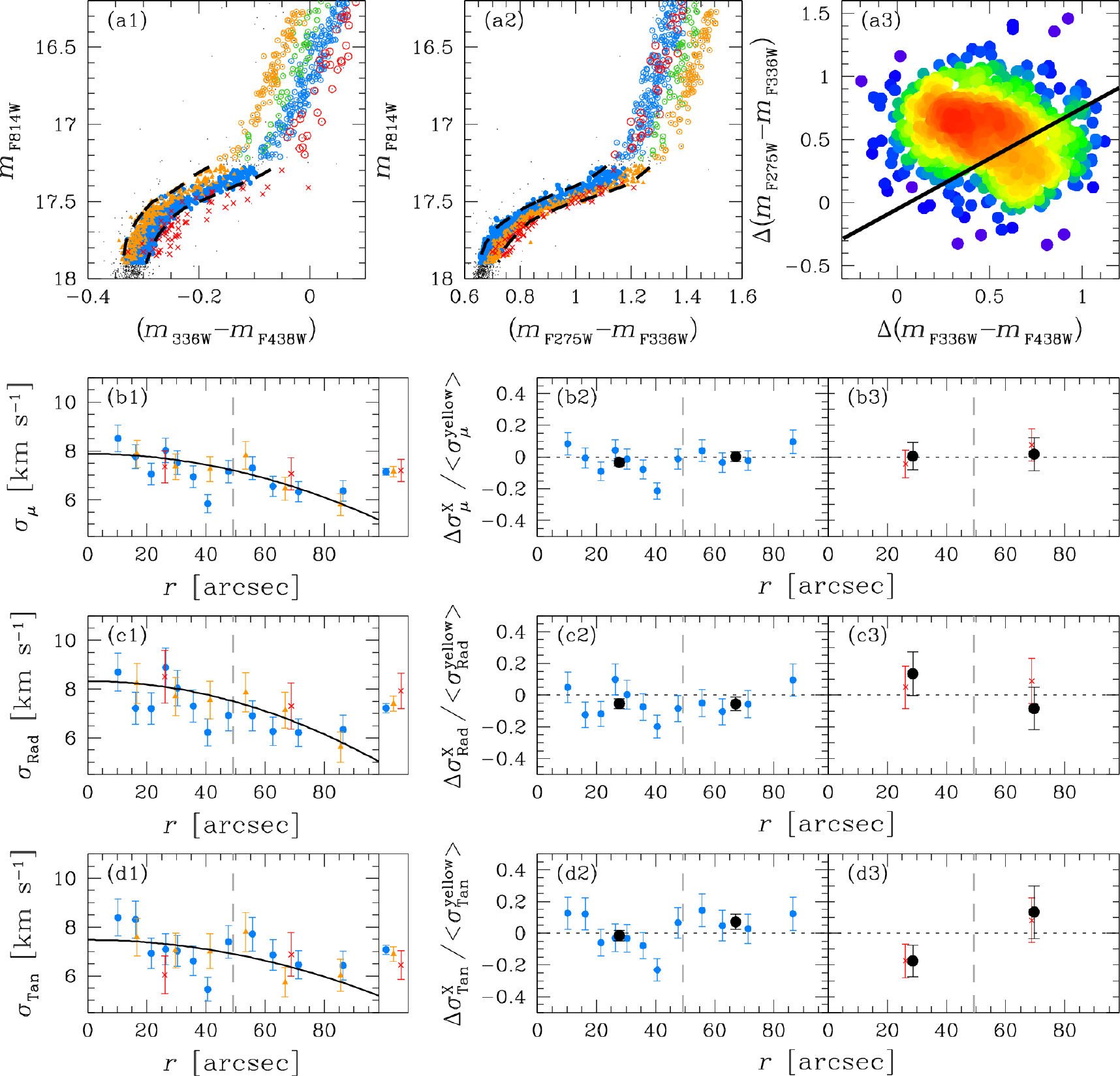}
  \caption{In panels (a1) and (a2) we present the $\eqmagi$ versus
    $(\eqmagu-\eqmagb)$ and the $\eqmagi$ versus $(\eqmaguv-\eqmagu)$
    CMDs of NGC~362, respectively. Open circles in panels (a1) and
    (a2) are RGB stars and are color-coded as in
    Fig.~\ref{fig:rgbpaper1}. The black, dashed lines are the fiducial
    lines adopted to rectify the SGBs analogous to the RGB study shown
    in Fig.~\ref{fig:rgbpaper1}. In panel (a3) we show the Hess
    diagram (in logarithmic scale) of the $\Delta (\eqmaguv-\eqmagu)$
    versus $\Delta (\eqmagu-\eqmagb)$ chromosome map of the SGB stars
    of NGC~362. We excluded from this map the red-SGB stars. Stars
    above this line are shown as azure dots in the previous panels,
    while the remaining stars are plot with yellow triangles. Red
    crosses represent red-SGB stars. Panels (b), (c), and (d) are
    similar to those in Fig.~\ref{fig:rgbpaper1} but for the mPOPs
    hosted in the SGB. See the text for details.}
  \label{fig:sgb3poppaper1}
\end{figure*}

In panels (a1) and (a2) of Fig.~\ref{fig:sgb3poppaper1}, we illustrate
the construction of the SGB chromosome map. First, we rectified the
SGBs in the $\eqmagi$ versus $(\eqmagu-\eqmagb)$ and $\eqmagi$ versus
$(\eqmaguv-\eqmagu)$ CMDs (panels a1 and a2) using the same method as
in the RGB analysis. The red crosses in these panels represent the red
SGB previously identified.

Then, we built the $\Delta (\eqmaguv-\eqmagu)$ versus $\Delta
(\eqmagu-\eqmagb)$ chromosome map. The Hess diagram of this chromosome
map is presented in panel (a3). For the sake of clarity, we excluded
from the Hess diagram of the chromosome map the red-SGB stars. The
remaining stars in the plot can be tentatively separated in a main
group and a tail by means of the black line in panel (a3). Hereafter,
we refer to the stars above and below the black line as azure- and
yellow-SGB stars, respectively.

In panels (a1) and (a2), we plot with azure dots and yellow triangles
the stars in each of the two groups identified in the more-populated
SGB, respectively. These two groups of SGB stars present a color
inversion between \magi versus $(\eqmagu-\eqmagb)$ and \magi versus
$(\eqmaguv-\eqmagu)$ CMDs (panels a1 and a2), similar to the behavior
of the mPOPs along the RGB. Furthermore, the yellow triangles are in
agreement with the 1G RGB stars, suggesting that they might be a 1G
population. The azure dots appear to be consistent with the 2G RGB
populations B and C, thus implying that they might be a 2G population.

\begin{figure*}
  \centering
  \includegraphics[width=0.8\textwidth]{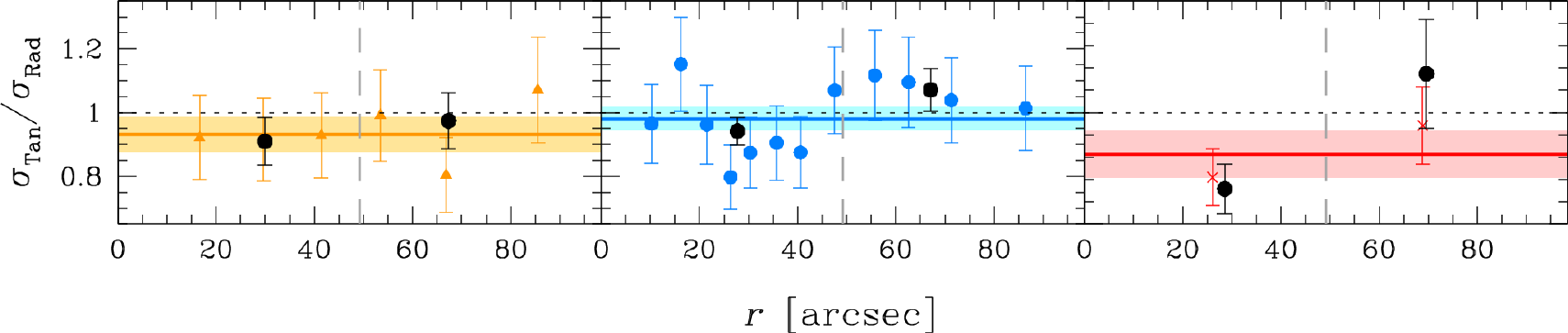}
  \caption{Similar to Fig.~\ref{fig:rgbpaper2}, but for the mPOPs on
    the SGB of NGC~362.}
  \label{fig:sgb3poppaper2}
\end{figure*}

In Fig.~\ref{fig:sgb3poppaper1}, we present $\sigma_\mu$ (panels b),
$\sigma_{\rm Rad}$ (panels c), and $\sigma_{\rm Tan}$ (panels c) of
SGB stars as a function of the radial distance. The velocity
dispersion in each bin in the plot is computed by considering 62, 50,
and 30 stars for the azure, yellow, and red SGBs, respectively. At the
$1\sigma$ level, the three SGB populations present the same
kinematics.

In Fig.\ref{fig:sgb3poppaper2} we show the tangential-to-radial
anisotropy profile of the SGB stars. All SGB populations are found to
be isotropic with the exception of the red-SGB stars, for which the
profile is significantly radially anisotropic at the $3\sigma$ level
for $r < r_{\rm h}$.

%%%%%%%%%%%%%%
\subsubsection{Main sequence}\label{ms}
%%%%%%%%%%%%%%

The analysis of the RGB and SGB stars is limited by small number
statistics. In this section we focus our attention on the more
plentiful MS stars.

\begin{figure*}
  \centering
  \includegraphics[width=0.8\textwidth]{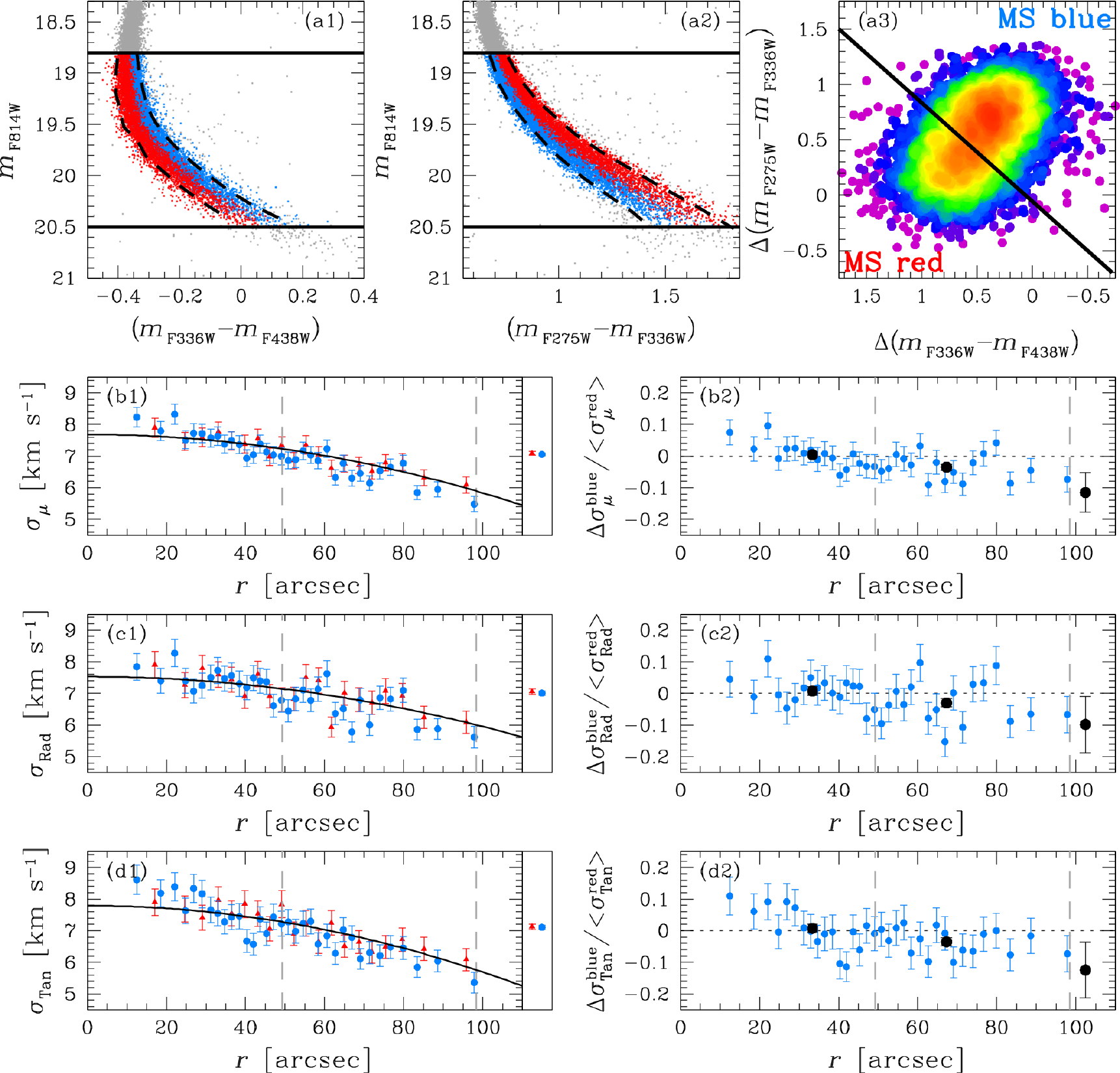}
  \caption{Similar to Fig.~\ref{fig:sgb3poppaper1} but for MS
    stars. The velocity dispersion in each bin in of panels (b), (c),
    and (d) is computed by considering 180 MS stars (except the
    outermost bin that is computed with 156 and 167 stars for MS red
    and blue, respectively). See the text for details.}
  \label{fig:mspaper1}
\end{figure*}

The mPOP tagging was performed as for the SGB stars using a UV-filter
chromosome map, which shows an elongated distribution of stars. From
the Hess diagram of the chromosome map (panel a3 of
Fig.~\ref{fig:mspaper1}), we found that the MS stars can be split into
a main group (hereafter ``MS blue'', in analogy with other studies of
MS mPOPs) and a tail (``MS red''). These two groups of stars are
clearly separated in the \magi versus $(\eqmagu-\eqmagb)$ and \magi
versus $(\eqmaguv-\eqmagu)$ CMDs (panels a1 and a2), while in colors
based in at least one optical filter such separation is less
evident. We considered only well-measured stars\footnote{In addition
  to the photometric selections described in Appendix~\ref{corr}, we
  discarded all stars with $|{\tt RADXS}| > 0.025$ in F275W, F336W,
  F438W, F606W and F814W filters.}  with $18.8 < \eqmagi < 20.5$
because in this magnitude interval the separation between MS blue and
red in the Hess diagram is clearer.

Stars that are O-rich/Na-poor (1G) are bluer than those O-poor/Na-rich
(2G) in $(\eqmagu-\eqmagb)$ color, but redder in $(\eqmaguv-\eqmagu)$
color \citep{2015AJ....149...91P}. As such, MS-blue stars are likely
2G stars, while MS-red objects belong to the 1G population. These
considerations are in agreement with the photometric and chemical
tagging performed for the RGB stars in Sect.~\ref{rgb}.

For $r < r_{\rm h}$, both MSs share the same velocity-dispersion
profile (panels b, c, and d in Fig.~\ref{fig:mspaper1}). For $r_{\rm
  h} < r < 2r_{\rm h}$, the MS-blue stars appear to be radially (at
the $2.4\sigma$ level) and tangentially (at the $2.7\sigma$ level)
colder than the MS-red stars. Finally, both MSs also have the same
kinematic profiles at the $1\sigma$ level for $r > 2r_{\rm h}$.

\begin{figure*}
  \centering
  \includegraphics[width=0.8\textwidth]{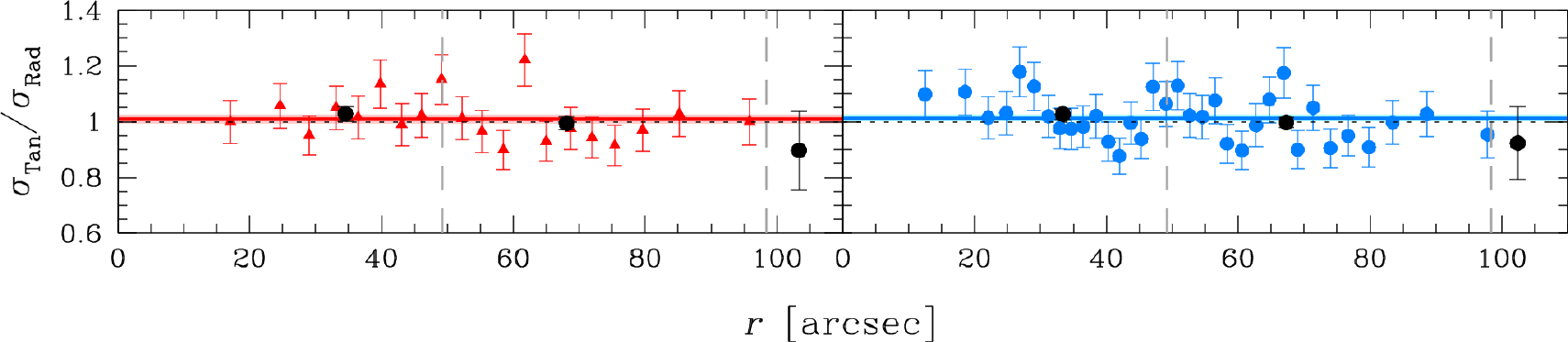}
  \caption{Similar to Figs.~\ref{fig:rgbpaper2} and
    \ref{fig:sgb3poppaper2} but for MS stars.}
  \label{fig:mspaper2}
\end{figure*}

The tangential-to-radial anisotropy (Fig.~\ref{fig:mspaper2}) shows
that the two groups are isotropic within our FoV.

%%%%%%%%%%%
\subsection{Energy equipartition}\label{eeq}
%%%%%%%%%%%

From numerical simulations, \citet{2013MNRAS.435.3272T} and
\citet{2016MNRAS.458.3644B} showed that the energy-equipartition state
($\sigma_\mu \propto m^{-\eta}$ with $\eta = 0.5$ and $m$ the stellar
mass) cannot be achieved, on account of the Spitzer
instability. \citet{2013MNRAS.435.3272T} showed that the stellar
velocity dispersion reaches a maximum value of $\eta \sim 0.15$ in the
core and then evolves, as in the other part of the cluster, to $\eta
\sim 0.08$.

We investigated the status of energy equipartition of NGC~362 as
follows. First, we divided the MS in 10 equally-populated bins of 2583
stars between $19 < \eqmagv < 24$. We considered only well-measured
stars (both astrometrically and photometrically) in F606W and F814W
filters.

\begin{figure*}
  \centering
  \includegraphics[width=\textwidth]{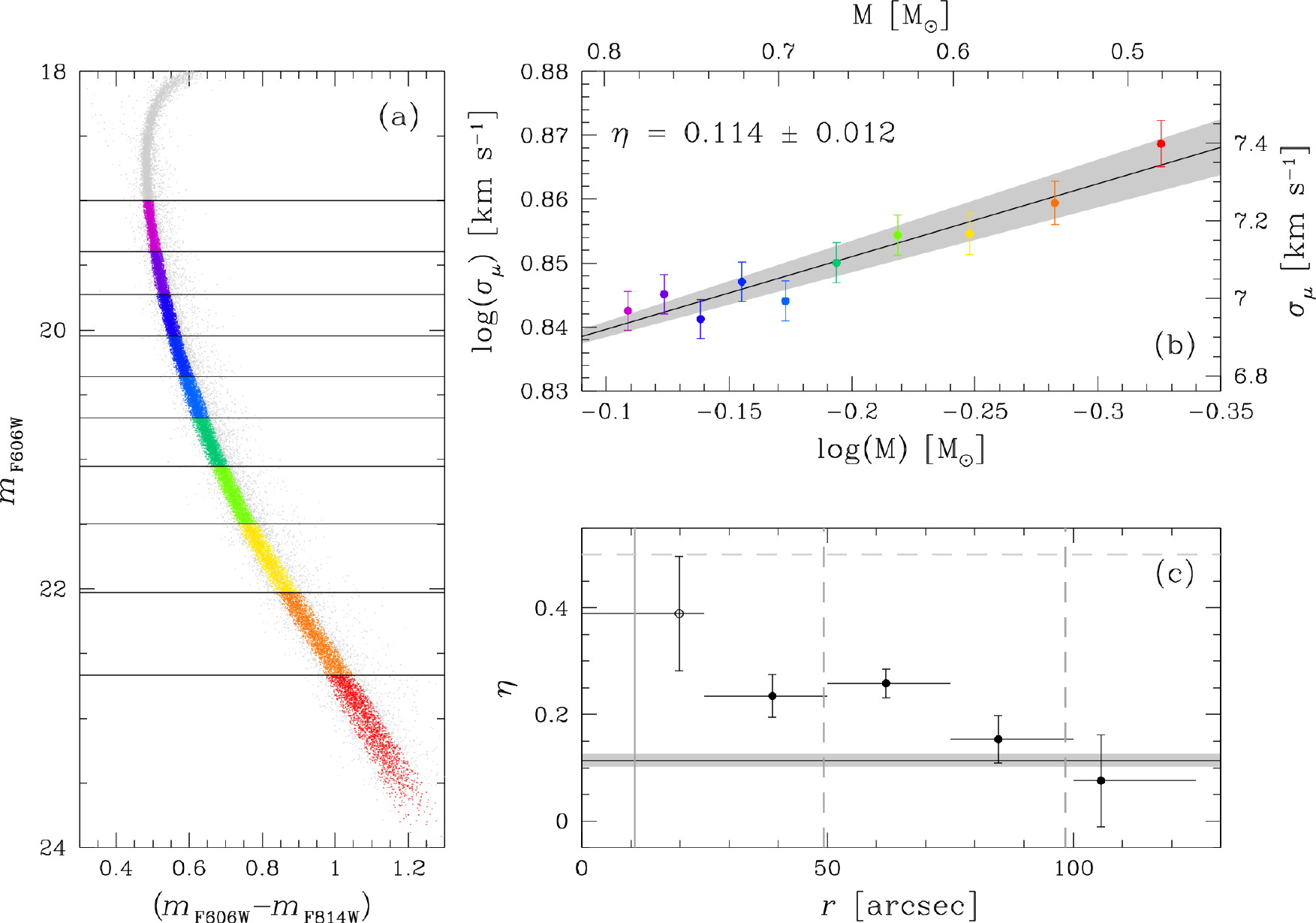}
  \caption{(a): \magv versus \colvi CMD of the MS of NGC~362 in which
    we defined 10 (equally-populated) bins color-coded from purple to
    red. Stars in these bins are used to compute the global level of
    equipartition of the cluster. The gray, horizontal lines define
    the limits of each magnitude bin. (b): combined velocity
    dispersion as a function of stellar mass. Both axes in the plot
    are in logarithmic scale. Points are color-coded as on the CMD in
    panel (a). The black line is the best fit straight line fit to the
    data. The slope of this line gives the level of energy
    equipartition $\eta$. (c): variation of $\eta$ as a function of
    radial distance. Each point is the local value of $\eta$ computed
    in steps of 25 arcsec from the center. Filled dots represent
    measurements in which stars adopted for the computation span a
    mass interval of $\sim 0.3 M_\sun$, while open circle indicates
    the case of $\Delta M \sim 0.2 M_\sun$. The horizontal error bars
    represent the radial interval of each point. The gray, solid
    vertical line is set at $r_{\rm c}$, while the two dashed lines
    are placed at $r_{\rm h}$ and $2r_{\rm h}$, respectively. The
    gray, dashed horizontal line refers to $\eta = 0.5$.}
  \label{fig:gomasstot1}
\end{figure*}

We computed the velocity dispersion $\sigma_\mu$ and the median
magnitude of the stars in each bin. Median magnitudes were transformed
into masses using a Darthmouth isochrone \citep{2008ApJS..178...89D}
with $\lbrack \rm Fe/H \rbrack = -1.26$, ${\rm E}(B-V) = 0.06$,
primordial He abundance and age of 11.46 Gyr \citep[as in,
  e.g.,][]{2017MNRAS.468.1038W}. Finally, we fitted the $\sigma_\mu$
versus mass values in a log-log plane with a weighted least-squares
straight line. The energy-equipartition parameter $\eta$ is the slope
of this straight line. The result is summarized in panels (a) and (b)
of Fig.~\ref{fig:gomasstot1}. We find:
\begin{equation}
  \eta = 0.114 \pm 0.012 \, .
\end{equation}
The half-mass relaxation time $t_{\rm rh}$ of a GC changes over time,
either increasing or decreasing in response to the cluster's dynamical
evolution, but its changes are expected to be small
\citep{1987degc.book.....S}. If we assume that the initial $t_{\rm
  rh}$, $t_{\rm rh}(0)$, is equal to $\sim 0.8$ Gyr as inferred by
\citet{2012A&A...539A..65Z}, we find $t_{\rm NGC~362} \sim 14 t_{\rm
  rh}(0)$. As such, our value of $\eta$ is in agreement with what is
expected from Fig.~6 of \citet{2013MNRAS.435.3272T}.

Our estimate of the state of energy equipartition is obtained by
considering all stars in the field, from the center to beyond 2$r_{\rm
  h}$. However, the level of energy equipartition in a GC is not the
same at all distances. We also investigated the local level of energy
equipartition in NGC~362 by dividing the sample into 5 radial bins of
25 arcsec each. The result is presented in panel (c).

The innermost interval presents $\eta \sim 0.4$, in contrast with the
\citet{2013MNRAS.435.3272T} $\eta_{\rm max}$ of $\sim 0.2$. However,
our estimate of $\eta$ in the innermost radial bin was obtained by
using stars covering a smaller mass range ($\Delta M \sim 0.2 M_\sun$)
than in all other bins in Fig.~\ref{fig:gomasstot1} ($\Delta M \sim
0.3 M_\sun$), and might have been overestimated because of a poor
straight-line fit.

The remaining points in panel (c) reveal that the level of energy
equipartition decreases from $\sim 0.25$ to $\sim 0.08$ as the radial
distance increases. This behavior is what we would expect as a result
of the dynamical evolution of the cluster
\citep{2017MNRAS.464.1977W}. The centermost regions of a GC are
expected to be the first to relax.

As shown in panel (c), the global value of $\eta$ (black horizontal
line) is not the average of the other points in the plot. The
difference between $\sigma_\mu$ for a high-mass star at the center and
at the outskirts of our field is smaller than for a low-mass star
because of the combined effects of energy equipartition and
hydrostatic equilibrium (see Fig.~\ref{fig:any2rmpaper}). As a
consequence, we expect the local $\eta$ at the center to be higher
than at the edge of the FoV because the low-mass stars at the center
are kinematically hotter. Since the most of the faint stars are
located in the outskirts of the FoV, the global value of $\eta$ is
closer to the outermost than the innermost local values of $\eta$.

~\newline
%%%%%%%%%%%%%%
\subsubsection{To collapse or not collapse}
%%%%%%%%%%%%%%

Previous studies of NGC 362 have not been able to conclusively
determine whether this cluster in the post-core-collapsed
phase. \citet{1989ApJ...339..904C} and \citet{1995AJ....109..218T} fit
the surface brightness profile with a King model with concentration
equal to 1.75 and 1.94, respectively, and classified NGC~362 as a
possible post-core collapse cluster. \citet{2013ApJ...778..135D}
showed that the observed star-count profile of NGC~362 is well
reproduced by either a mild power law ($\alpha \sim -0.2$) or a double
King profile \citep[similar to what was found in the
  post-core-collapsed GC NGC~6752 by][]{2003ApJ...595..179F}, another
indication that some dynamical processes have occurred in the cluster
core. The advanced dynamical state of this cluster was also pointed
out by \citet{2012Natur.492..393F} from the analysis of the BS radial
distribution in the context of the so-called ``dynamical
clock''. Furthermore, the presence of two sequences of BSs in NGC~362
(see Sect.~\ref{twobss}) and their radial distributions also seem to
support the post-core-collapsed scenario
\citep{2009Natur.462.1028F,2013ApJ...778..135D}. However, none of
these studies was able to clearly infer the pre- or
post-core-collapsed state of NGC~362.

Recently, \citet{2018MNRAS.tmpL..16B} proposed the kinematic
concentration $c_k$ as a diagnostic for core collapse, based entirely
on the internal kinematics of a cluster. This parameter is defined as:
\begin{equation}
  c_k = \frac{m_{\rm eq}(r<r_{50})}{m_{\rm eq}(r_{50})} \, ,
\end{equation}
where $m_{\rm eq}(r<r_{50})$ and $m_{\rm eq}(r_{50})$ are the mass
scale parameters measured by considering all stars within the 50\%
Lagrangian radius and between the 40\% and 60\% Lagrangian radii,
respectively. The mass scale parameter $m_{\rm eq}$ quantifies the
level of energy equipartition of a cluster
\citep{2016MNRAS.458.3644B}. In detail, \citet{2016MNRAS.458.3644B}
describe the relation between the velocity dispersion $\sigma_\mu$ and
the stellar mass as:
\begin{equation}\label{equation3}
  \sigma (m) = \left\{
  \begin{array}{lc}
      \sigma_0 \exp(-\frac{1}{2}\frac{m}{m_{\rm eq}}) & {\rm if}~{m \le m_{\rm eq}} \\
      \sigma_0 \exp(-\frac{1}{2}) (\frac{m}{m_{\rm eq}})^{-\frac{1}{2}} & {\rm if}~{m > m_{\rm eq}} \\
  \end{array}
  \right. \, ,
\end{equation}
where $\sigma_0$ is the velocity dispersion for $m = 0$. The relation
between $\sigma_\mu$ and mass describes the different behavior of the
energy equipartition for high- and low-mass stars. The cut-off at $m =
m_{\rm eq}$ avoids unphysical values of $\eta$. The relation between
$\eta$ and $m_{\rm eq}$ is shown in Eq. 4 of
\citet{2016MNRAS.458.3644B}.

We estimated the local ($m_{\rm eq}(r_{50})$) and global ($m_{\rm
  eq}(r<r_{50})$) level of energy equipartition as follows. The
velocity dispersions were computed as described in
Sect.~\ref{eeq}. Instead of the 50\% Lagrangian radius, we adopted the
half-light radius as it is a direct observable. This assumption is
justified since the results obtained with either the 40\%, 50\% or
60\% Lagrangian radii are similar \citep{2018MNRAS.tmpL..16B}. For
$m_{\rm eq}(r_{50})$, we considered the stars within $\pm 0.5r_{\rm
  c}$ from $r_{\rm h}$. Then, we performed a weighted least-square fit
to the data using Eq.~\ref{equation3} with $\sigma_0$ and $m_{\rm eq}$
as free parameters. We find:
\begin{equation}
  \left\{
  \begin{array}{lcl}
    m_{\rm eq}(r<r_{50}) & = & 1.60 \pm 0.32 \, M_\sun \\
    m_{\rm eq}(r_{50})   & = & 1.03 \pm 0.18 \, M_\sun \\
  \end{array}
  \right. \, ,
\end{equation}
which for NGC~362 implies:
\begin{equation}
  c_k = 1.54 \pm 0.41 \, .
\end{equation}
According to \citet{2018MNRAS.tmpL..16B}, a cluster with $c_k > 1$ has
reached the core-collapsed state. Our result is the first inference of
the post-core-collapsed state of NGC~362 based solely on
kinematics. For the sake of completeness, we also computed the value
of $c_k$ by adopting the half-mass radius given in
\citet{2012A&A...539A..65Z} and found $c_k = 1.31 \pm 0.35$, further
confirming NGC~362 to be a post-core-collapse GC.

The value of $m_{\rm eq}(r<r_{50})$ is also qualitatively in agreement
with the predictions of Fig.~6 of \citet{2016MNRAS.458.3644B},
although their simulations did not involve core-collapse GCs. In
particular, \citet{2016MNRAS.458.3644B} provide a relation between
$m_{\rm eq}(r<r_{50}=r_{\rm h})$ and $n_{\rm rel} = t_{\rm age} /
t_{\rm rc}$, the ratio between the cluster age and the core relaxation
time. For $t_{\rm age} = 11.46$ Gyr \citep{2017MNRAS.468.1038W} and
$t_{\rm rc} = 0.06$ Gyr \citep[][2010 edition]{1996AJ....112.1487H},
we obtain from their Eq.~6 $m_{\rm eq}(r<r_{50}) \sim 1.60$, in
excellent agreement with our estimate.

%%%%%%%%%%%
\subsection{Blue stragglers}\label{bss}
%%%%%%%%%%%

In \citetalias{2016ApJ...827...12B}, we measured cluster kinematics to
estimate the masses of BSs in a number of GCs. We repeat this analysis
here using our updated catalog.

BSs are stars bluer and brighter than the MS turn-off, that is, they
apparently sit on the MS where stars more massive than those at the
turn-off would be found. It is thought that they are still present on
the MS and have not evolved off as expected for stars of that mass
because they gained their mass only recently, most likely as a result
of stellar collisions or binary evolution
\citep[e.g.,][]{2009Natur.457..288K}. As GCs are collisional systems,
the stars interact and share energy. As a result, the clusters exhibit
some degree of energy equipartition whereby more massive stars move
more slowly than lower mass stars. If the BSs are indeed more massive
than the RGB stars then it follows that we should expect the BSs to be
moving more slowly than the RGB stars. It is this that allows us to
use kinematics to estimate the mass of a BS population in a
cluster. This approach is similar to that described by
\citet{2016ApJ...830..139P}, where they compared the radial
distributions of BSs with those of other stellar populations to infer
the mass of the BSs in 47\,Tuc.

\begin{figure*}
  \centering
  \includegraphics[width=0.8\textwidth]{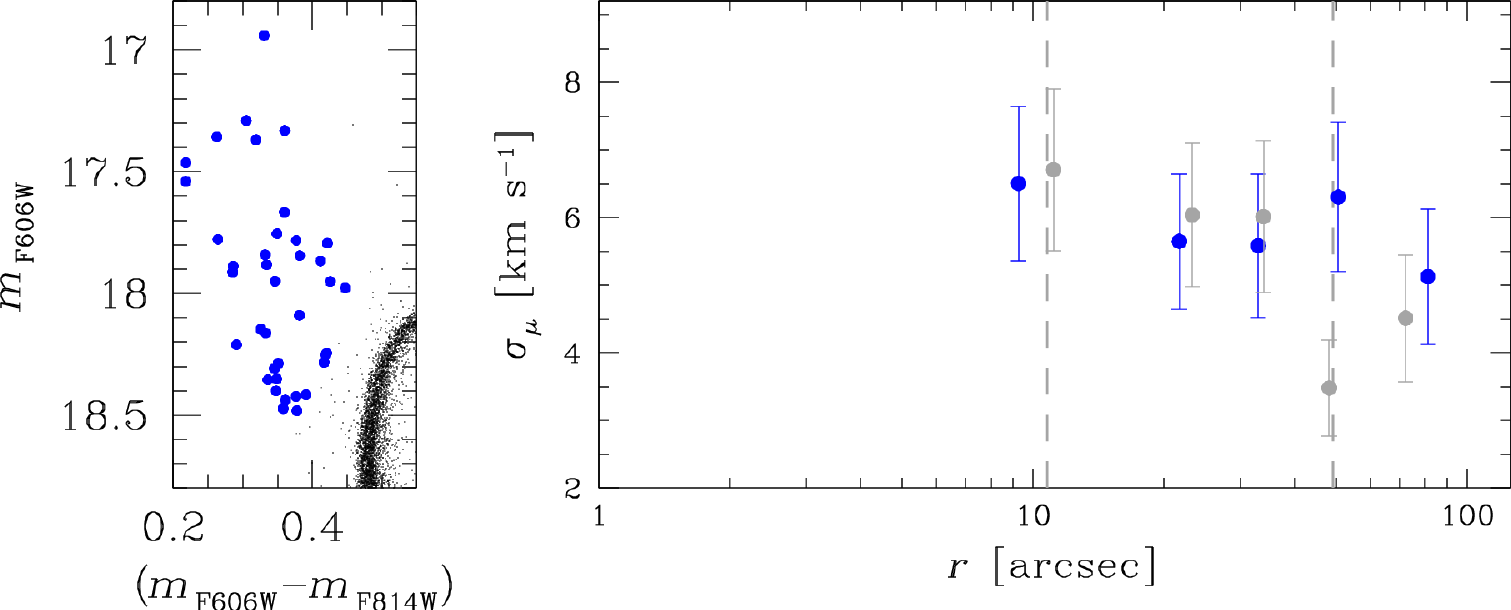}
  \caption{In the \magv versus \colvi CMD on the left, we selected a
    sample of BSs (blue dots) that are used to compute the
    $\sigma_\mu$ radial profile on the right panel. Gray dots are the
    $\sigma_\mu$ values obtained in \citetalias{2016ApJ...827...12B}
    using old PMs.}
  \label{fig:gobss}
\end{figure*}

To do this, we estimate the velocity dispersion profiles of the BSs
and the RGB stars from the kinematic data. First, we selected BSs from
the \magv vs. $(\eqmagv-\eqmagi)$ CMD and computed their $\sigma_\mu$
as a function of radius in 5 equally-populated bins each containing 8
stars. The result is shown in Fig.~\ref{fig:gobss}. Blue dots
represent the $\sigma_\mu$ values of BSs computed with our new PMs,
the gray dots are from \citetalias{2016ApJ...827...12B}. The two
trends are in good agreement.

Then, we selected all well-measured NGC~362 members with $\eqmagv <
19.8$ (i.e., one magnitude below the MS turn-off). We computed the
combined velocity-dispersion profile in (i) one bin with only stars
within 2 arcsec from the center of NGC~362, (ii) five
equally-populated bins from 2 arcsec out to $r_{\rm c}$, (iii) 5 bins
from $r_{\rm c}$ to $r_{\rm h}$, and (iv) 10 bins from $r_{\rm h}$ to
the edge of the FoV.

We assume the velocity profiles of the BSs and the RGB stars have the
same shape but that the BS dispersions are a fraction $\alpha$ lower
owing to their mass difference, that is $\sigma_\mathrm{\textsc{BS}} =
\alpha \sigma_\mathrm{\textsc{RGB}}$.

In our previous work, this was a two-step process. In
\citetalias{2015ApJ...803...29W}, we measured the PM dispersion from
NGC\,362 from the catalog in \citetalias{2014ApJ...797..115B} and to
this we fit a monotonically-decreasing 4$^{\rm th}$-order polynomial
that was forced to be flat at the center. In
\citetalias{2016ApJ...827...12B}, we measured the dispersion profile
for the BSs, used the best-fitting RGB polynomial as our fiducial
profile, and estimated the factor $\alpha$ that provided the best fit
to the BS profile. Here we are able to combine both steps together,
and fit for the polynomial parameters and factor $\alpha$
simultaneously. We also fit the polynomial directly to the individual
stars and not to the dispersion profile. We use the affine-invariant
Markov-Chain Monte Carlo (MCMC) method \textsc{emcee}
\citep{2013PASP..125..306F} to find and sample the region of parameter
space that best fits the data. We extract 10\,000 points from the
final MCMC parameter distributions (100 points at each of 100 steps,
selected at 10-step intervals) to use as our fit sample.

\begin{figure}
  \centering
  \includegraphics[width=\columnwidth]{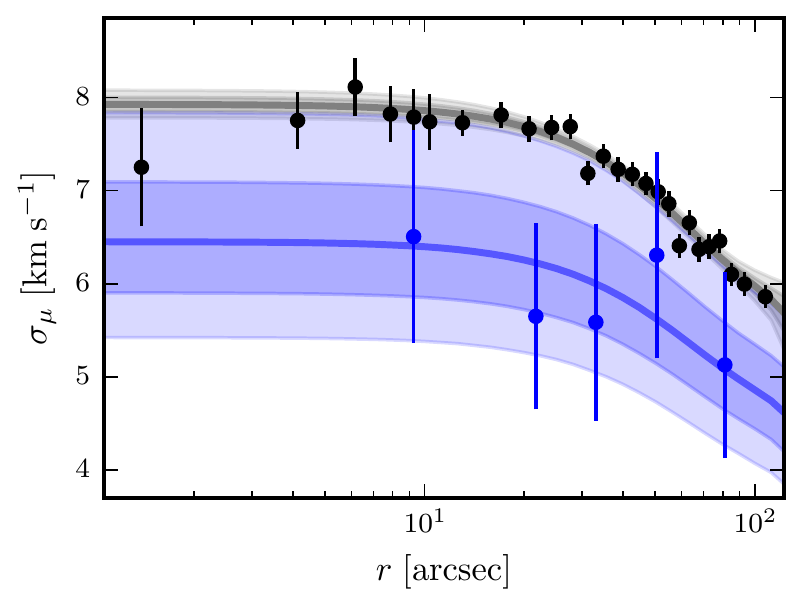}
  \caption{Measured velocity dispersion profiles for the RGB stars
    (black points) and the BSs (blue points). To the RGB stars, we fit
    a monotonically-decreasing 4$^{\rm th}$-order polynomial with a
    flat center; we assume that that BS profile is the same shape but
    rescaled by a factor $\alpha$. The black and blue lines show the
    median of the resulting fits, and the dark and light shaded
    regions show the 1$\sigma$ and 2$\sigma$ percentile regions.}
  \label{fig:fit_dispersion}
\end{figure}

Figure~\ref{fig:fit_dispersion} shows the best-fitting polynomials,
along with the dispersion profiles previously calculated. Note, the
dispersion profiles are shown for visualization purposes, the fits
were performed on individual stars. The black points show the measured
dispersion profile for the RGB stars and the blue points show the
measured dispersion profile for the BSs as in Fig.~\ref{fig:gobss}.

For each point in our final fit sample, we calculate the predicted RGB
and BS dispersion profiles. The gray and blue lines show the medians
of those profiles for the RGB stars and BSs, respectively. The
dark-gray and dark-blue areas show the region spanned by 1$\sigma$
(15.9 and 84.1) percentiles of the fitted profiles, and the light-gray
and light-blue areas are show the region spanned by the 2$\sigma$ (2.3
and 97.7) percentiles of the fitted profiles. From these fits, we find
$\alpha = 0.81^{+0.08}_{-0.07}$, where the best-estimate is the median
of the points that constitute our fit sample, and the uncertainties
are estimated using the 1$\sigma$ percentiles.

In order to turn this velocity difference into a mass difference, we
need to relate the scale factor $\alpha$ of the velocity dispersion
profiles with scale factor $f$ of the masses, that is
$M_\mathrm{\textsc{BS}} = f M_\mathrm{\textsc{RGB}}$. If we assume
that the cluster is in partial equipartition then $\sigma \propto
M^{-\eta}$, and it follows that $f = \alpha^{-\frac{1}{\eta}}$. In
\citetalias{2016ApJ...827...12B}, we had no estimate for $\eta$
directly, so we estimated $\eta$ from simulations by
\citet{2016MNRAS.458.3644B} and assumed a boxcar distribution centered
on $\eta$ and with half-width $\eta/3$ in order to propagate our
uncertainty in this value. We now have an estimate (and uncertainty)
for $\eta$ at the average distance of the BSs ($r \sim 40$ arcsec):
$\eta = 0.23 \pm 0.04$ (see the bottom-right panel of
Fig.~\ref{fig:gomasstot1} in Sect.~\ref{eeq}). We choose to use this
value here, but otherwise adopt a similar procedure as before and draw
$\eta$ from a boxcar centered on 0.23 and with half-width 0.04. We use
the fit sample to calculate a mass ratio $f = 2.45^{+1.24}_{-0.83}$,
again taking the median and 1$\sigma$ percentiles of the resulting
distribution.

Finally, we can estimate the mass of the BS stars. In
\citetalias{2016ApJ...827...12B}, we used isochrone fitting to
estimate the mass of stars at the MS turn-off and assumed that the RGB
stars have the same mass, thus $M_\mathrm{\textsc{RGB}} = 0.80 \,
M_\sun$. Combining this value with our estimate of mass fraction $f$,
we obtain a final mass estimate for the BSs of $M_\mathrm{\textsc{BS}}
= 1.96^{+0.99}_{-0.66} \, M_\sun$. This is smaller than our previous
estimate $2.18^{+0.73}_{-0.42} \, M_\sun$, but in agreement, within
the uncertainties.

This value of the BS mass remains the largest mass estimate among
those provided in \citetalias{2016ApJ...827...12B}, and is somewhat
offset from the mean value found for all the clusters of
$\overline{M}_\mathrm{\textsc{BS}} = 1.22 \pm 0.12 \, M_\sun$. A
possible explanation of the discrepancy might be the incomplete sample
of BSs analyzed, in particular close to the center of
NGC~362. \citet[][see Fig.~9]{2013ApJ...778..135D} found that the
median distances of red- and blue-sequence BSs in NGC~362 are 15 and
28 arcsec, respectively. If we use the value of $\eta$ in
Fig.~\ref{fig:gomasstot1} at these distances ($\eta = 0.39 \pm 0.11$)
and assume that our BS sample is representative of all BSs in NGC~362,
the mass of the BS decreases to $1.36^{+0.40}_{-0.30} \, M_\sun$, in
good agreement with the overall mean value found previously.

%%%%%%%%%%%%%%
\subsubsection{The two BS populations of NGC~362}\label{twobss}
%%%%%%%%%%%%%%

NGC~362 is known to host two sequences of BSs
\citep{2013ApJ...778..135D}. The red sequence should be constituted by
BSs resulting from mass-transfer binaries, the blue sequence from
collisions. We studied the global kinematic behavior of these stars as
follows.

\begin{figure}
  \centering
  \includegraphics[width=\columnwidth]{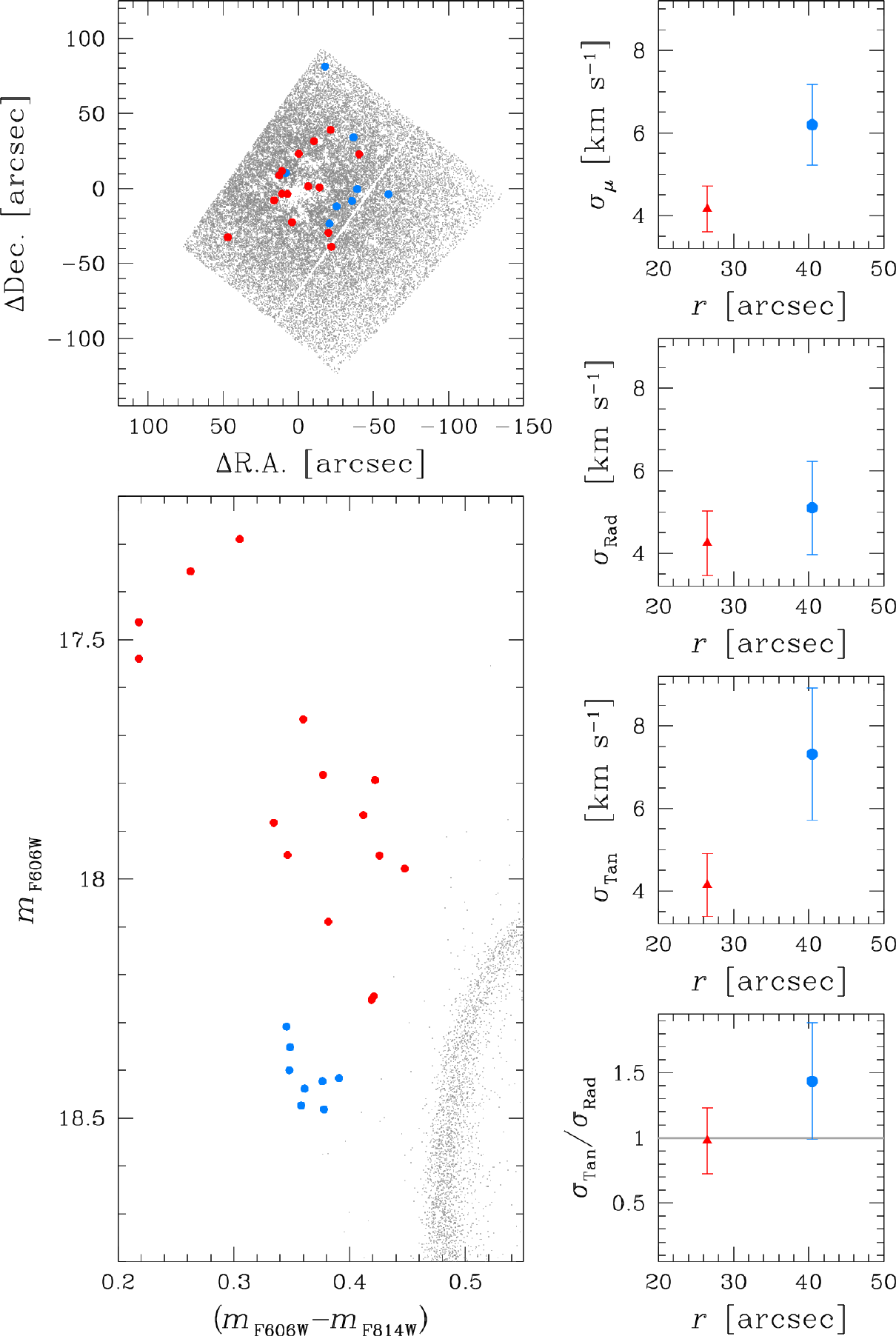}
  \caption{Kinematic analysis of the two BS groups in NGC~362 (as
    defined on the \magv versus \colvi CMD of the bottom-left
    panel). We find the red BSs to be more centrally concentrated than
    the blue BSs (top-left panel), which was also reported by
    \citet{2013ApJ...778..135D}. In the right-hand panels, we show
    (from top to bottom) the combined velocity dispersion
    ($\sigma_\mu$), the radial-velocity dispersion ($\sigma_{\rm
      Rad}$), the tangential-velocity dispersion ($\sigma_{\rm Tan}$)
    and the tangential-to-radial anisotropy radial profiles.}
  \label{fig:gobssgrp2}
\end{figure}

We defined the two samples of BSs as shown in Fig.~7 of
\citet{2013ApJ...778..135D}, i.e., by defining the fiducial loci of
the two groups of BSs in the $m_{\rm F555W}$ versus $(m_{\rm
  F555W}-m_{\rm F814W})$ CMD. The well-measured\footnote{The sample of
  well-measured BSs defined in Sect.~\ref{bss} was refined by applying
  the photometric selections described in Appendix~\ref{corr} to the
  WFC3/UVIS filters F555W and F814W used for the BS tagging.} BSs for
the kinematic analysis are shown in the CMD in
Fig.~\ref{fig:gobssgrp2}. For each of the two BS populations, we
computed the value of $\sigma_\mu$, $\sigma_{\rm Rad}$, $\sigma_{\rm
  Tan}$, and the tangential-to-radial anisotropy in one radial bin
covering the entire FoV (right panels in Fig.~\ref{fig:gobssgrp2}).

The red and blue BSs share the same kinematics at the 1$\sigma$
level. There is only a negligible difference in the tangential
component $\sigma_{\rm Tan}$ between red and blue BSs that goes in the
direction that we would expect from their formation mechanisms.

Finally, the red BSs are found to be more concentrated than the blue
BSs, confirming the results of \citet{2013ApJ...778..135D}.

%%%%%%%%%%%
\subsection{Cluster rotation}\label{rot}
%%%%%%%%%%%

To date, only a handful of PM-based studies of the rotation of GCs in
the plane of the sky have been performed
\citep{2000A&A...360..472V,2003AJ....126..772A,2013ApJ...779...81M,2017ApJ...844..167B,2017ApJ...850..186H,2018ApJ...853...86B}.

Regarding our target, \citet{2017MNRAS.469.4740J} concluded from a
spectroscopic analysis that the rotation of NGC~362 is small, if
present at all. More recently, \citet{2018MNRAS.473.5591K} found a
maximum rotation of $\sim 2.3$ km s$^{-1}$ at the center, decreasing
to 0.6 km s$^{-1}$ at a distance of 1 arcmin. Here, we measured the
amount of rotation of NGC~362 in three different ways.

Our PMs are computed by adopting the cluster stars as reference, and
as such any direct trace of rotation in the reference population is
canceled out by our linear transformations. However, any sign of
rotation would be transferred (with the opposite sign) to
background/foreground objects in the field.

\begin{figure*}
  \centering
  \includegraphics[width=\textwidth]{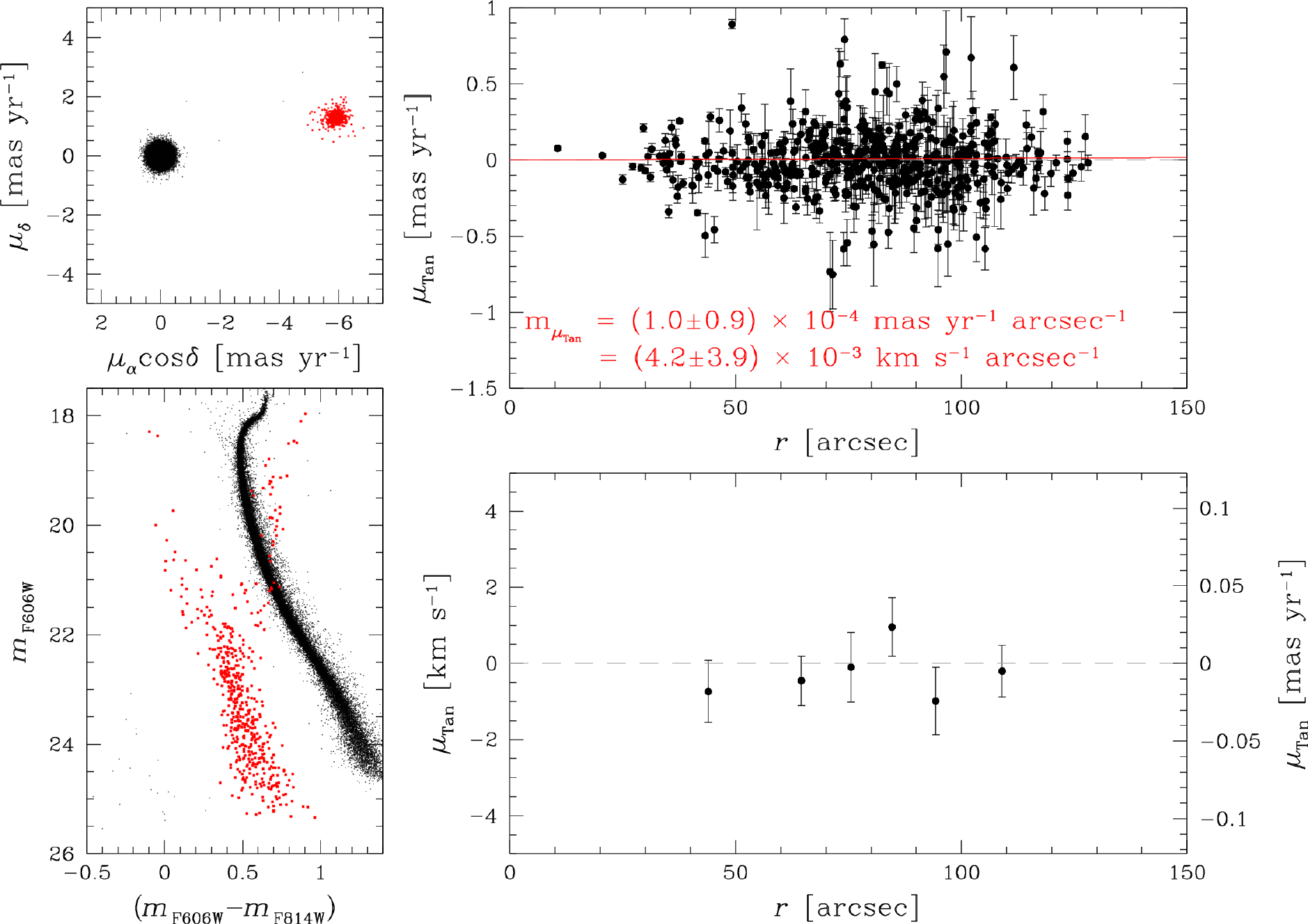}
  \caption{(Top-left): VPD of relative PMs. Members of NGC~362 are
    centered on the origin (black dots), while SMC stars have a
    distribution centered around $(-5.9, 1.3)$ mas yr$^{-1}$ (red
    points). (Bottom-left): \magv versus \colvi CMD of NGC~362 (black)
    and SMC (red) stars. (Bottom-right): $\mu_{\rm Tan}$ of SMC stars
    as a function of distance from the center of NGC~362. (Top-right):
    tangential component of the PMs ($\mu_{\rm Tan}$) of SMC stars as
    a function of distance from the center. The gray, dashed
    horizontal line is set at 0 mas yr$^{-1}$. The red line is a
    straight-line fit to the points. Both panels on the right indicate
    that NGC~362 is not rotating in the plane of the sky.}
  \label{fig:rot4}
\end{figure*}

NGC~362 is located in front of the SMC. As such, we used the SMC
background stars to probe the rotation of NGC~362 in the plane of the
sky as was done by \citet{2017ApJ...844..167B} for 47\,Tuc. To exclude
outliers from our data, we selected only stars with a PM error lower
than 1 mas yr$^{-1}$ and in which the PM fit was performed with at
least 50 images, but we did not apply any other PM-based selection.

We selected the SMC stars in the VPD (top-left panel of
Fig.~\ref{fig:rot4}), divided them in 6 equally-populated radial bins
(85 stars per bin), and measured the tangential component of the PM
$\mu_{\rm Tan}$ in each bin. We find no evidence of plane-of-the-sky
rotation for NGC~362 (bottom-right panel of Fig.~\ref{fig:rot4}).

We also used an alternative method to infer the presence of rotation
as was done by \citet{2013ApJ...779...81M}. We fit a straight line to
the tangential component of the PM of the SMC stars as a function of
distance from the center of NGC~362. The slope of the straight line is
consistent with 0, meaning that the cluster is not rotating.

Finally, we used a third method, starting with the assumptions made by
\citet{2017ApJ...850..186H} in their PM analysis of the GC
47\,Tuc. The authors found that the PMs of 47\,Tuc stars are skewed in
the tangential direction, implying that the cluster is rotating in the
plane of the sky. The rotating nature of 47\,Tuc had been previously
inferred and characterized by \citet{2017ApJ...844..167B} with a
multi-field analysis similar to that we used for NGC~362. This
skew-based method is of particular interest when no other reference
systems than the cluster itself are available. More recently, the same
technique was adopted to infer the presence of differential rotation
in the mPOPs of the GC $\omega$\,Cen \citep{2018ApJ...853...86B}.

Following \citet{2018ApJ...853...86B}, we computed the amount of skew
in our PMs by measuring (i) the skewness value $G_1$ and the
corresponding significance test $Z_{G_1}$ \citep{Cramer1997}, and (ii)
the third-order Gauss-Hermite moment $h_3$
\citep[e.g.,][]{1993ApJ...407..525V}. In general, a symmetric
distribution has $-0.5 < G_1 < 0.5$ \citep{Bulmer1979}. The
significance level of this value is given by $Z_{G_1}$: if $|Z_{G_1}|
> 2$, the result is significant at a $>2\sigma$ level, otherwise no
conclusion can be inferred. The significance level of $h_3$ is instead
given by its error.

For NGC~362 we find:
\begin{equation}
  \left\{
  \begin{array}{l}
    G_1 = -0.01 \\ Z_{G_1} = -0.71 \\ h_3 = -0.006 \pm 0.004 \\
  \end{array}
  \right. \, .
\end{equation}

The values of $G_1$ and $Z_{G_1}$ suggest that the PM distribution is
not skewed at the $3\sigma$ level. The third-order Gauss-Hermite
moment $h_3$ provides further support to the conclusion that NGC~362
has no internal rotation.

%%%%%%%%%%%
\subsection{Absolute PM}\label{apm}
%%%%%%%%%%%

We visually inspected the \textit{HST} images of the cluster and
identified two background galaxies, which we used to compute the
absolute PM of NGC~362.  The PMs of the two galaxies did not pass our
astrometric quality-selection criteria (on account of their
non-point-source nature). However, we are interested in high accuracy
and not necessary in high precision to infer the absolute PM of the
cluster.

\begin{figure}
  \centering
  \includegraphics[width=1.0\columnwidth]{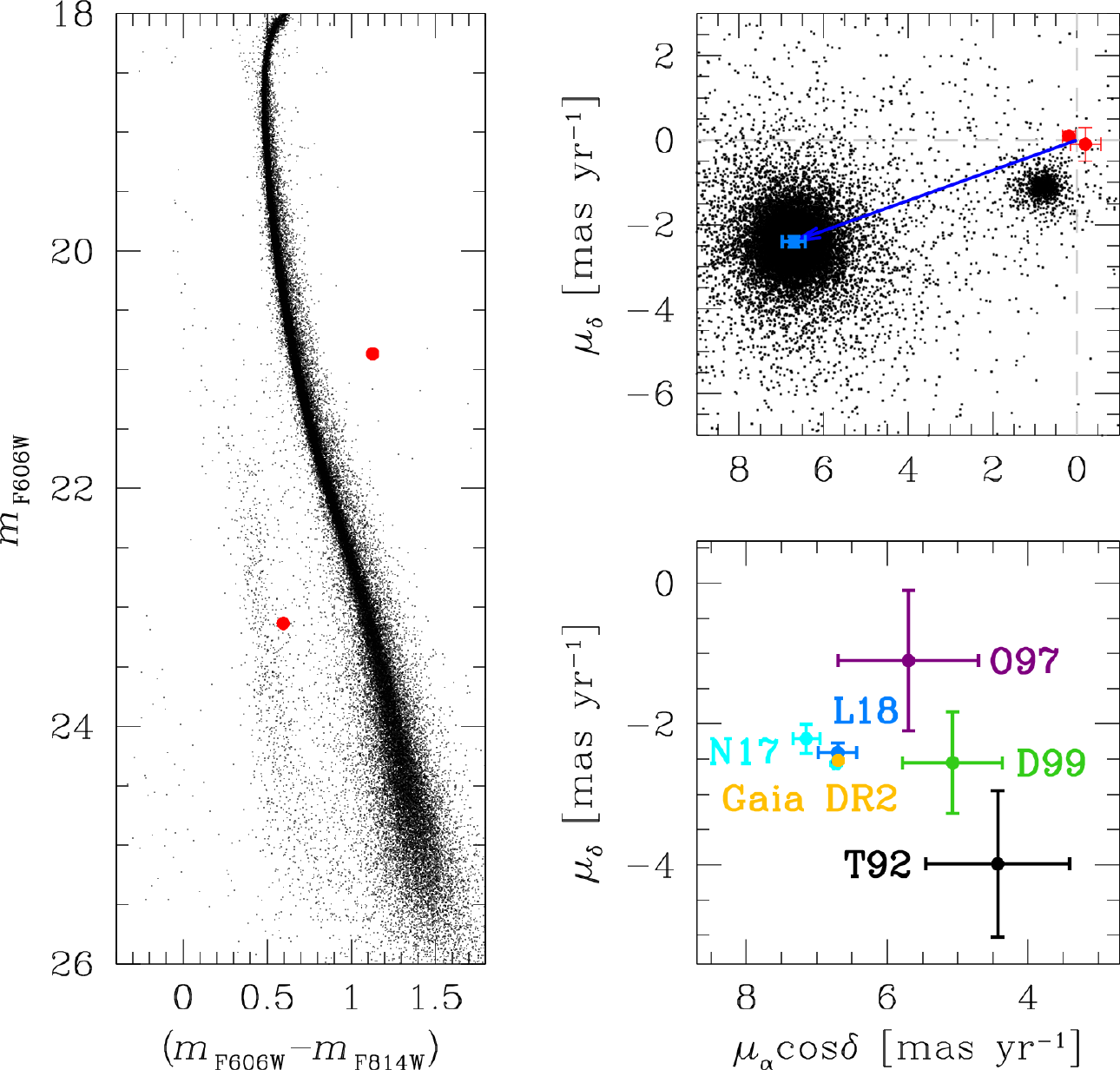}
  \caption{(Left): \magv versus \colvi CMD of NGC~362. The two
    galaxies used to compute the absolute PM of the cluster are shown
    as red dots. All other objects in the plot are depicted in
    black. (Top-right): VPD of the absolute PMs in equatorial
    coordinates. The red points (with error bars) are the PM of the
    galaxies. The blue line indicates the absolute motion of the
    cluster, the azure point with $1 \sigma$ uncertainty represents
    the absolute PM of NGC~362. (Bottom-right): Comparison of our
    estimate the absolute PM of NGC~362 (L18, azure) with the
    literature values: \citet[T92, black]{1992A&AS...93..293T},
    \citet[O97, purple]{1997NewA....2..477O}, \citet[D99,
      green]{1999AJ....117.1792D}, and \citet[N17,
      cyan]{2017MNRAS.471.1446N}. The estimate obtained with the PMs
    of the Gaia DR2 catalog is shown in yellow. The error bars of the
    Gaia-based PM have the same size of the point.}
  \label{fig:gal11}
\end{figure}

Since our PMs are computed relative to the bulk motion of NGC~362, the
background galaxies have a motion equal to the absolute PM of NGC~362,
but with opposite sign. We find (see Fig.~\ref{fig:gal11}):
\begin{equation}
  \begin{array}{ll}
    (\mu_\alpha\cos\delta,\mu_\delta)_{\rm NGC~362} \\ =
    (6.703\pm0.278,-2.407\pm0.135)~{\rm mas~yr^{-1}} \, .\\
  \end{array}
\end{equation}
The errors are the standard errors in the mean. It is straightforward
to measure the absolute PM of SMC stars in our field. We refer to
Appendix~\ref{smc} for the analysis of the PMs of the SMC stars.

The expected effect of parallax in our FoV given the temporal coverage
of our images is lower than 0.026 mas yr$^{-1}$ (see
Appendix~\ref{smc}). Therefore, we chose not to correct it because it
is within the PM uncertainties of the reference galaxies.

As discussed in, e.g., \citet{2018ApJ...854...45L}, the absolute PM
measured in a given field can be the combination of the amount of
rotation of the system and of the motion of the center of mass
(COM). Furthermore, projection effects arise when there are different
lines of sight between the COM and the analyzed field. All these
effects must be taken into account to infer the true absolute PM of
NGC~362. However, in our case we do not have to include any of these
contributions because (i) our field is centered on NGC~362, and (ii)
we did not measure any systemic rotation (Sect.~\ref{rot}).

In the top-right panel of Fig.~\ref{fig:gal11}, we present the VPD of
the absolute PM of stars in our field.  The azure point represents our
estimate of the absolute PM of NGC~362. In the bottom-right VPD we
zoom-in around the location of our estimate of the absolute PM of
NGC~362 and we compare it with literature values. We find an excellent
agreement with the most-recent PM values given by
\citet{2017MNRAS.471.1446N} and with the value inferred with the PMs
of the Gaia DR2 catalog by \citet{2018arXiv180409381G}.

%%%%%%%%
\section{Conclusions}
%%%%%%%%

In this sixth paper of the series, we present an improved version of
the data reduction aimed at better characterizing the PMs of stars in
crowded environments, and applied it to the GC NGC~362.

Thanks to the new reduction pipeline we were able to (i) increase the
number of detected sources and improve the overall PM astrometric
precision by a factor of 3-5 with respect to the previous version of
the PM catalog \citepalias{2014ApJ...797..115B}, particularly in the
innermost regions of the GC, (ii) reach a PM precision of $\sim 10$
$\mu$as yr$^{-1}$ for bright stars, and (iii) measure the PMs of faint
stars with high precision (e.g., a factor 20 better than the expected
end-of-mission precision of Gaia at its faint end $\eqmagv \sim
21.5$).

We separated the mPOPs along the RGB, SGB and MS of the cluster and
measured their velocity dispersions. All mPOPs show the same
kinematics. We find only a marginal signature of 2G RGB stars of
population B having a lower tangential velocity dispersion than 1G
stars (at the 2.2$\sigma$ level). This evidence is similar to what has
been recently found in other GCs and with numerical simulations
\citep{2013ApJ...771L..15R,2015ApJ...810L..13B,2018ApJ...853...86B},
but the difference between $\sigma_{\rm Rad}$ and $\sigma_{\rm Tan}$
is not large enough to also create a significant radial anisotropy as
in, e.g., NGC~2808 and $\omega$\,Cen.

We also studied the level of energy equipartition of the cluster and
its dependence on the radial distance from the cluster's center. Our
results show that the degree of equipartition is stronger at smaller
distances from the cluster's center in agreement with what expected
from the effects of two-body relaxation
\citep{2013MNRAS.435.3272T,2017MNRAS.464.1977W}. Furthermore, we
inferred that NGC~362 is in a post-core-collapsed state by comparing
the local and global levels of energy equipartition. The
classification of NGC~362 as a post-core-collapsed cluster is based
entirely on the internal kinematics of the cluster; previous studies
based the surface-brightness profile showed that this cluster could be
modeled by a high-concentration King model, but the studies were
unable to provide a firm conclusion concerning its pre- or
post-core-collapsed state. Our result provides an additional example
of the key role that the study of the internal kinematics can play in
building a complete dynamical picture of GCs.

We refined the estimate of the average dynamical mass of the BSs
hosted in this cluster, finding good agreement with the previous value
of the BS mass published in \citetalias{2016ApJ...827...12B}. NGC~362
is also known to host two sequences of BSs. We analyzed their
velocity-dispersion radial profiles and find no differences in the
kinematics.

We investigated the rotation in the plane of the sky of NGC~362 in
three different ways. We find no evidence of significant
plane-of-the-sky rotation.  Although projection effects might
contribute to hide the presence of rotation, this result suggests that
this cluster is in an advanced stage of its evolution and has lost
most its initial angular momentum \citep[see, e.g.,][for numerical
  studies illustrating the gradual loss of rotation during a GC's
  evolution]{1999MNRAS.302...81E,2007MNRAS.377..465E,2017MNRAS.469..683T}.

Finally, we used two galaxies in the background to measure the
absolute PM of NGC~362. As a by product of our investigation, we also
calculated the absolute PM of the SMC stars in our field.

The PM catalog of NGC~362 is made available to the community. The
description of the catalog is provided in Appendix~\ref{cat}.

%%%%%%%%%
\section*{Acknowledgments}
%%%%%%%%%

M.L. and A.B. acknowledge support from STScI grant GO
13297. M.L. acknowledges partial support by the Universit\`a degli
Studi di Padova Progetto di Ateneo CPDA141214 ``Towards understanding
complex star formation in Galactic globular clusters''. The authors
thank the anonymous Referee for the thoughtful suggestions that
improved the quality of the paper. Based on observations with the
NASA/ESA \textit{HST}, obtained at the Space Telescope Science
Institute, which is operated by AURA, Inc., under NASA contract NAS
5-26555. This work has made use of data from the European Space Agency
(ESA) mission {\it Gaia} (\url{https://www.cosmos.esa.int/gaia}),
processed by the {\it Gaia} Data Processing and Analysis Consortium
(DPAC,
\url{https://www.cosmos.esa.int/web/gaia/dpac/consortium}). Funding
for the DPAC has been provided by national institutions, in particular
the institutions participating in the {\it Gaia} Multilateral
Agreement.

\bibliographystyle{aasjournal}

\appendix

%%%%%%%%
\section{First- and second-pass photometry}\label{red}
%%%%%%%%

For the data reduction, we employed only \texttt{\_flt}-type
exposures, since they preserve the unresampled pixel data for
appropriate point-spread-function (PSF) fitting. ACS/WFC and WFC3/UVIS
\texttt{\_flc} images have been pipeline corrected for
charge-transfer-efficiency (CTE) defects as described in
\citet{2010PASP..122.1035A}. No CTE correction is available for
ACS/HRC data\footnote{We investigated the presence of CTE-related
  systematic effects in our PMs and found them to be negligible.}.

The first-pass photometry is performed with a single wave of finding,
and no neighbor subtraction is applied prior to the PSF fit.  We
extracted positions and fluxes of all detectable sources via empirical
PSF fitting. The PSF models of each exposure were obtained by
perturbing the
publicly-available\footnote{\href{http://www.stsci.edu/~jayander/STDPSFs/}{http://www.stsci.edu/$\sim$jayander/STDPSFs/}.}
library PSFs of each \textit{HST} camera/filter. We derived
spatially-variable perturbation PSFs for the ACS/WFC and WFC3/UVIS
images, while a single (spatially-constant) perturbation PSF model was
applied to ACS/HRC data.  Stellar positions were corrected for
geometric distortion by using the distortion solutions provided by
\citet{2004acs..rept....3A,2006acs..rept....1A},
\citet{2009PASP..121.1419B} and \citet{2011PASP..123..622B}.

Because it does no neighbor subtraction, the first-pass photometry is
severely limited in the centermost regions of the cluster, where there
is significant crowding. Our second-pass photometry is specifically
designed to address this issue.

Before running the second-pass photometry algorithm, we need to set-up
a common reference-frame system in which all stars can be measured
consistently. We adopted a reference system in which the $X$ and $Y$
axes point toward West and North, respectively, and the center of the
cluster \citep[provided by][]{2010AJ....140.1830G} is placed at
position $(5000,5000)$. The pixel scale of the master frame is set to
be exactly 40 mas\,pixel$^{-1}$, very similar to that of WFC3/UVIS,
and an intermediate value between those of ACS/HRC and ACS/WFC. We set
up orientation and scale of our reference system with the Gaia DR1
catalog. Then, we used only bright, unsaturated, relatively isolated
stars in common between our single-exposure catalogs and the Gaia DR1
to derive general, six-parameter linear transformations that were used
to iteratively cross-identify stars of each individual exposure on the
master-frame plane. Master-frame positions are then the average of
these transformed positions. For each filter, stellar magnitudes of
each exposure were zero-pointed to match those of the longest
available exposure before averaging.

The second-pass photometry is run using \texttt{KS2}, a sophisticated
\texttt{FORTRAN} routine based on the code developed to reduce the
``ACS Globular Cluster Treasury Survey'' data
(\citealt{2008AJ....135.2055A}). \texttt{KS2} starts from the outputs
of the first-pass photometry (PSFs and transformations) and
simultaneously reduces all individual epochs/filters/exposures at
once. Objects are measured in three different methods by
\texttt{KS2}. In this work we considered only the method-\#1
measurements, in which stellar positions and fluxes are obtained
through PSF fitting of the individual neighbor-subtracted exposures.
This method is best suited to high-precision PM analysis.

\texttt{KS2} allows us to select a subset of exposures to be used for
the finding stage. For this task we chose images taken with ACS/HRC
(F435W) and ACS/WFC (F606W $+$ F814W) for the following reasons.
First, ACS/HRC data are the most suitable to probe the very center of
this core-collapse GC (within its core radius $r_{\rm c} = 0.18$
arcmin), since they provide the highest angular resolution. Also, the
GO-10775 ACS/WFC data were taken at a somewhat intermediate epoch and
offer the largest overlap with other exposures. \texttt{KS2} measures
stars in single exposures if they are within a 1-pixel searching
radius from the position measured during the finding stage. Some
background objects, in particular SMC stars, have moved by more than 1
pixel between the first and the last available epoch, so choosing an
intermediate epoch for the finding process allowed us to find and
measure all sources.

\texttt{KS2} outputs several diagnostic parameters: \textsf{QFIT}
value\footnote{The \textsf{QFIT} represents the linear-correlation
  coefficient (similar to the Pearson coefficient) between the values
  of the real pixels and those of the PSF models. The closer to unity
  the \textsf{QFIT} is, the better the PSF fit.}, the magnitude rms,
the fractional flux due to neighbors within the fitting radius prior
to neighbor subtraction ($o$), the ratio between the number of
individual exposures used to measure a stellar position and flux and
the total number of exposures actually found for the star ($r_{\rm
  N}$), and the shape parameter ${\tt RADXS}$\footnote{The ${\tt
    RADXS}$ value is the excess/deficiency of flux outside of the
  fitting radius with respect to that predicted by the PSF. It is
  particularly effective is separating faint stars from artifacts and
  background galaxies \citep[see also][]{2008ApJ...678.1279B}.}. These
quality parameters will be used later in Sect.~\ref{analysis} to
select the best-measured sources.

Finally, the \texttt{KS2}-based magnitudes were calibrated in the
Vega-mag system following prescriptions given in
\citet{2017ApJ...842....6B}. Our photometry is corrected for
differential reddening as described in \citet{2012A&A...540A..16M} and
\citet{2017ApJ...842....7B}. We refer to \citet{2017ApJ...842....7B}
for the detailed description of the methodology.

%%%%%%%%
\section{Relative proper motions}\label{rpm}
%%%%%%%%

As a by product, \texttt{KS2} also provides neighbor-subtracted
stellar positions and fluxes in the raw reference system of each
exposure (hereafter, the raw catalogs):\ a clear advantage over the
catalogs produced by the first-pass photometry. PMs are computed using
these catalogs. We considered only stars measured in at least both
F606W and F814W filters\footnote{We have exposures in F814W filter
  obtained with either the ACS/WFC or the WFC3/UVIS cameras. In the
  paper, we always refer to the ACS/WFC F814W filter as simply F814W
  filter unless explicitly declared otherwise.}, and we excluded all
catalogs based on the F275W filter, because of color-dependent
systematic effects in the F275W geometric-distortion correction
\citep{2011PASP..123..622B}.

\citetalias{2014ApJ...797..115B} developed iterative procedures to
compute high-precision stellar PMs. These procedures have been
recently improved and discussed by \citet{2018ApJ...853...86B}, and
represent the state-of-the-art in observations with \textit{HST}. In
the following, we describe the outline of the PM computation and
highlight the few but significant changes we applied to our data set.

Each iteration starts by cross-identifying stars in each raw catalog
with those measured by \texttt{KS2} on the master frame, PM-shifted at
the epoch of the raw catalog, by means of general, six-parameter
linear transformations of a set of reference bright, unsaturated
cluster members. At the first iteration, stellar PMs are assumed to be
zero, and membership is determined solely on the basis of stellar
locations on the CMD. Since cluster members define the
transformations, the computed PMs will be \textit{relative} to the
bulk motion of the cluster.

For a given star, its master-frame transformed positions as a function
of epoch are fit with a least-squares straight line, the slope of
which is a direct estimate of the star's PM. This fitting procedure is
itself iterated, and involves data rejection and sigma clipping. We
refer the reader to \citetalias{2014ApJ...797..115B} for more
details. The last least-squares fit is performed with
locally-transformed master-frame stellar positions, based on the
closest 45 reference stars, as was done in
\citet{2018ApJ...853...86B}. Local transformations help in mitigating
large-scale systematic residuals.

At the end of each iteration, master-frame positions can be adjusted
to match the epoch of each observation to minimize mismatches during
the cross-identification step, as described in
\citet{2018ApJ...853...86B}. The PM-computation converges when there
are at most negligible differences between master-frame positions from
one iteration to the next.

%%%%%%%%%%%
\subsection{Correction of the PM systematics}\label{corr}
%%%%%%%%%%%

\citet{2018ApJ...853...86B} computed the PMs of $\omega$\,Cen using
the same techniques described in this paper. In their analysis, they
pointed out that the PMs were affected by low- and high-frequency
systematic effects. The former are related to the different overlaps
between the data sets, and hence with the temporal baseline. The
latter are fine-scale structures due to uncorrected CTE and distortion
residuals.

\begin{figure*}
  \centering \includegraphics[width=\textwidth]{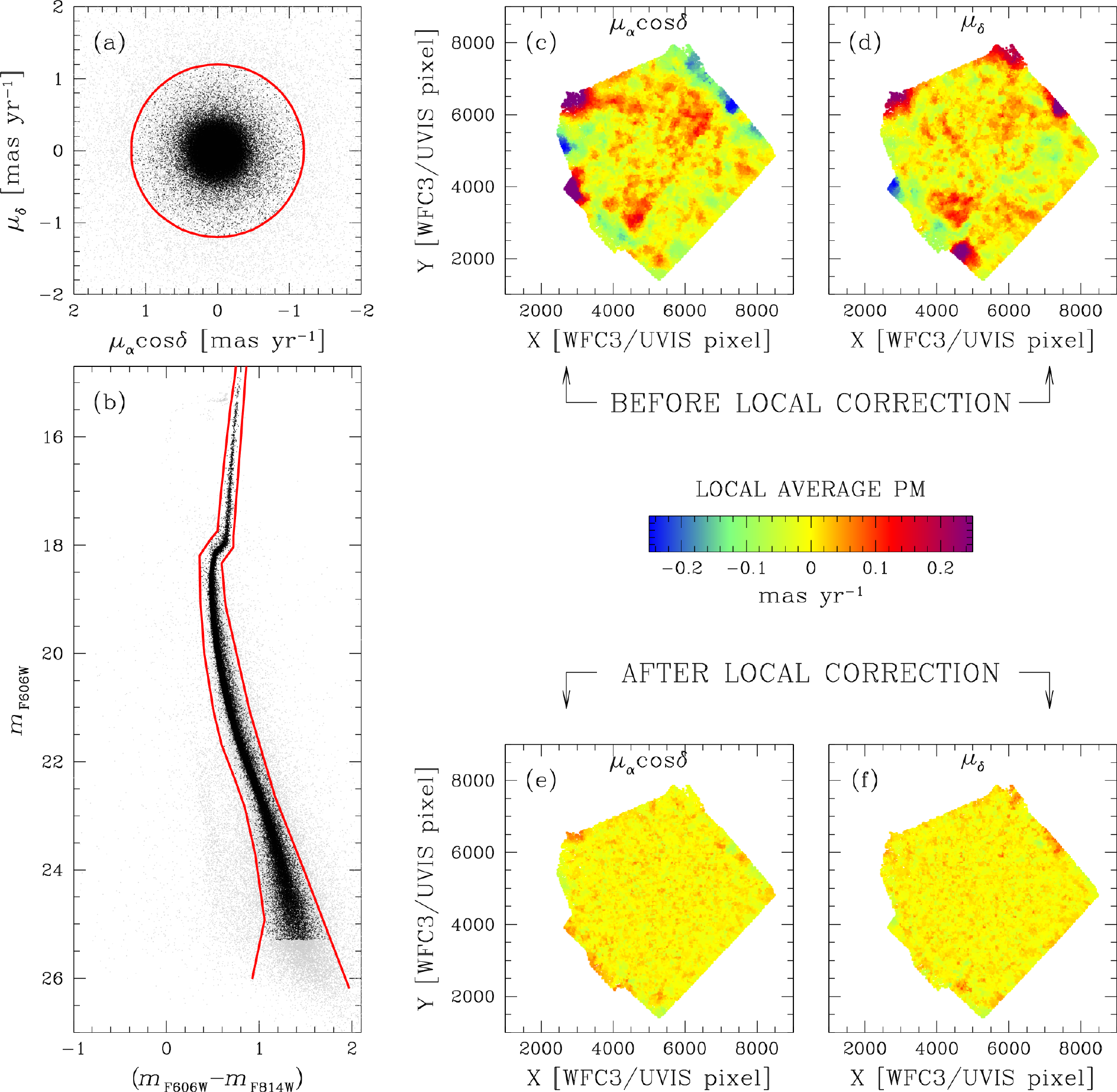}
  \caption{Overview of the correction for the high-frequency PM
    systematic effects. Well-measured cluster members are selected
    according to their PM in the VPD (within the red circle in panel
    a) and their location on the \magv versus \colvi CMD (within the
    two red lines in panel b). In panels (c) and (d) we show the maps
    of the local PM along $\alpha \cos\delta$ and $\delta$ (panel d)
    before the high-frequency correction. Each point is color-coded
    according to the color bar in the middle. The corresponding maps
    based on corrected PM are shown in panels (e) and (f) for
    $\mu_\alpha \cos\delta$ and $\mu_\delta$, respectively.}
  \label{fig:plotgo1}
\end{figure*}

First, we considered only stars with (i) \textsf{QFIT} larger than the
85$^{\rm th}$ percentile at any given magnitude\footnote{All objects
  with a \textsf{QFIT} larger than 0.99 are also included. Sources
  with a \textsf{QFIT} smaller than 0.6 are always discarded. A
  similar procedure was applied to the magnitude rms: stars with rms
  smaller than 0.01 mag are always retained, and all objects with a
  magnitude rms larger than 0.25 mag are rejected.}, (ii) magnitude
rms lower than the 85$^{\rm th}$ percentile, again at any given
magnitude, (iii) flux greater than 2.5 times the local rms of the sky,
(iv) $o$ parameter smaller than 1, and (v) $r_{\rm N}$ ratio greater
than 0.5. We also excluded all stars fainter than $\eqmagv \sim 25.3$
(F606W instrumental magnitude $\sim -6.5$). Finally, bonafide cluster
members are defined as well-measured stars with a PM smaller than 1.2
mas yr$^{-1}$ (panel a in Fig.~\ref{fig:plotgo1}) and lying within the
two ridge lines (drawn by hand) in the \magv versus \colvi CMD of
panel (b).

Low-frequency systematic effects reveal themselves as shifts of the
bulk motion of cluster stars from the origin of the VPD. To correct
for these low-frequency systematic effects, we first divided selected
stars into different groups according to the temporal baseline used to
compute their PMs. For each group, we then computed the
$3\sigma$-clipped median PM and corrected the PM of the stars
accordingly.

The maps of the local PM components along $\alpha \cos\delta$ and
$\delta$ (panels c and d) reveal the presence of uncorrected
high-frequency systematic effects with amplitude as large as $\pm 0.2$
mas yr$^{-1}$. Each point in the maps is color-coded according to the
median PM of the closest 100 reference stars.  In order to remove
these effects, we performed an a-posteriori correction as described in
\citetalias{2014ApJ...797..115B}. Briefly, for each object we selected
the closest 100 cluster stars. These stars were used to compute the
high-frequency correction defined as the median value of their PM in
each coordinate. In panels (e) and (f) we show the local PM maps after
the high-frequency effects were corrected.

As discussed in \citet{2018ApJ...853...86B}, at each step of the
correction we included in the error budget the standard error on the
median of the correction by summing it in quadrature with the PM
errors, thus artificially increasing the intrinsic velocity dispersion
of the cluster stars. This is not a problem for our studies of the
internal kinematics of NGC~362 because we account for the PM
uncertainties when we measure the observed PM dispersion. In any case,
our PM catalog contains both the original (uncorrected) and the
corrected PMs with the corresponding errors to allow users to choose
the best option for their investigation.

%%%%%%%%%%%
\subsection{Comparison with the PM catalog of Paper~I}\label{comp}
%%%%%%%%%%%

\begin{figure}
  \centering \includegraphics[width=\columnwidth]{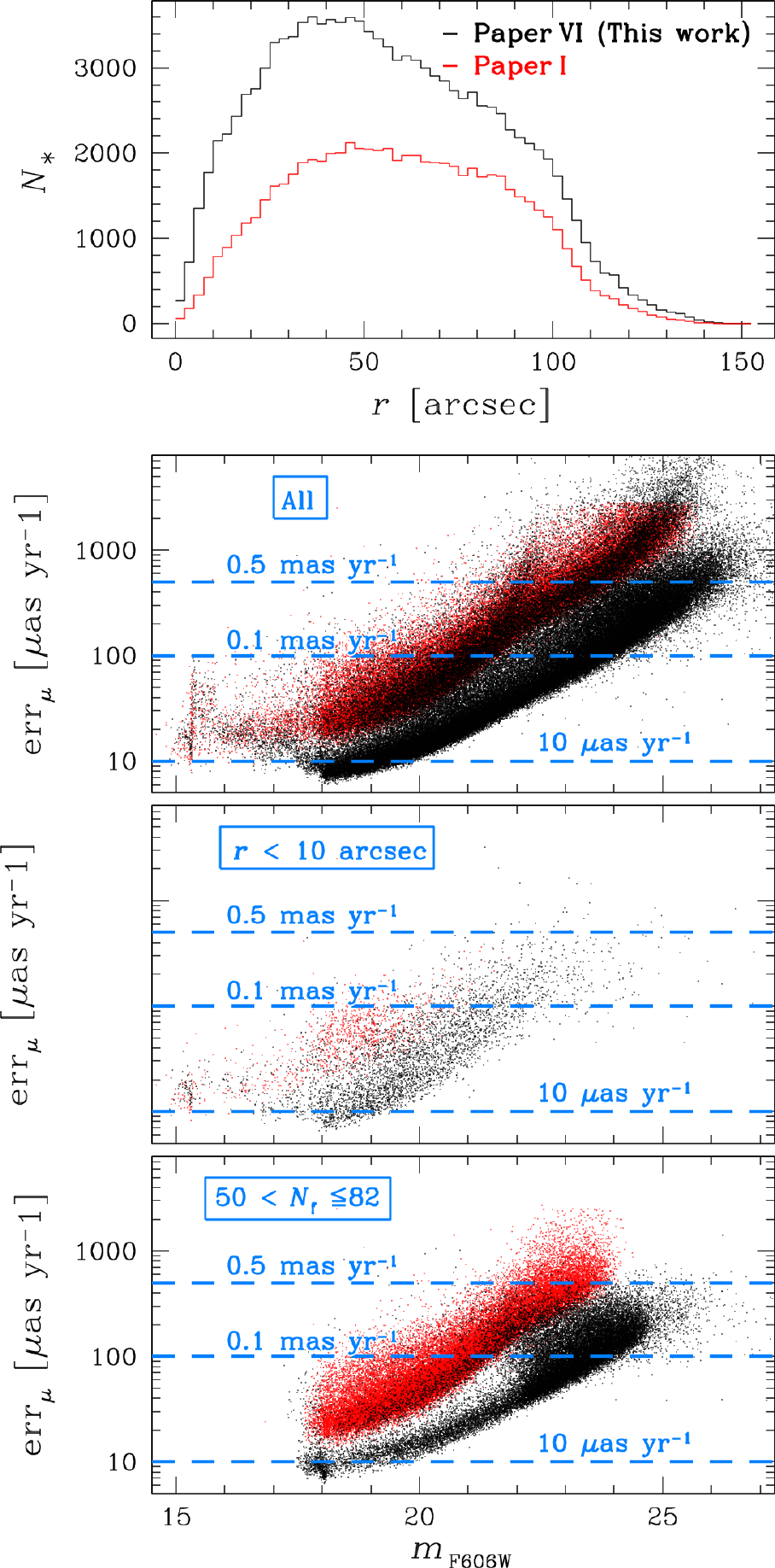}
  \caption{Comparison between the PMs obtained in this paper
    (Paper~VI, in black) and in \citetalias{2014ApJ...797..115B} (in
    red). The two histograms in the top panel show the number of
    sources as a function of distance from the cluster
    center. Overall, our new catalog contains more sources. In the
    three remaining panels, we compare the PM errors. From top to
    bottom, we show all stars, stars within 10 arcsec from the center
    of NGC~362, and stars with PM obtained by using between 50 and 82
    exposures, respectively.}
  \label{fig:plot2bellini}
\end{figure}

To better highlight the differences with the previous data reduction,
we made a comparison between the old and new PMs. The result is
presented in Fig~\ref{fig:plot2bellini}.

In our analysis we considered all objects with a PM measurement and,
for a fair comparison, we compared only the uncorrected PMs, since the
two papers performed different PM corrections. In the top panel, we
show a histogram of the objects in the catalog as a function of the
distance from the center of NGC~362. The new catalog has an overall
higher number of objects compared to that of
\citetalias{2014ApJ...797..115B}, in particular in the innermost 10
arcsec.

The astrometric precision reached in \citetalias{2014ApJ...797..115B}
and in our paper for well-measured stars brighter than the SGB level
is of about 30 $\mu$as yr$^{-1}$. For stars at $\eqmagv \sim 18$, the
PM precision achieved in our new catalog is typically three times
better than that of \citetalias{2014ApJ...797..115B}. At the faint end
of the \citetalias{2014ApJ...797..115B} catalog ($\eqmagv \sim 25$),
the median err$_\mu$ value is $\sim 1.6$ mas yr$^{-1}$, while in our
new catalog the PMs at this magnitude level have a median error of
$\sim 0.3$ mas yr$^{-1}$, a factor five better.

Second-pass photometry is more effective than first-pass photometry in
very crowded regions as the cluster's center because of the
neighbor-subtraction stage. As such, we compared the PM errors of
stars within 10 arcsec from the center of NGC~362 (second panel from
the bottom in Fig.~\ref{fig:plot2bellini}). The astrometric precision
reached for bright stars at the SGB level very close to the center of
NGC~362 is of the order of 10 $\mu$as yr$^{-1}$ in our new catalog,
and of $\sim 40$ $\mu$as yr$^{-1}$ in the old catalog, a factor 4
worse. For stars at the HB level ($\eqmagv \sim 15.3$), the value of
err$_\mu$ is comparable.

In the bottom panel of Fig.~\ref{fig:plot2bellini} we compared only
stars measured by using between 50 and 82 images (the maximum number
of overlapping exposures of \citetalias{2014ApJ...797..115B}). The new
PM errors in our paper are again a factor $\sim 4$ better than those
in the old catalog, and the PM-error distribution looks overall
tighter than that of the old catalog, meaning that stars at the same
magnitude are measured with about the same precision, regardless of
the crowding.

These examples summarize some of the differences between first- and
second-pass photometry. While the position of very-bright stars can be
measured reasonably well even in very-crowded environments without
neighbor subtraction, the fainter the stars and the more crowded the
field, the more important the neighbor subtraction
becomes. Furthermore, \texttt{KS2} allows us to find and measure stars
as faint as $\eqmagv \sim 27$, two magnitudes fainter than in the
old-PM catalog.

%%%%%%%%
\section{SMC}\label{smc}
%%%%%%%%

\begin{figure}
  \centering
  \includegraphics[width=.7\columnwidth]{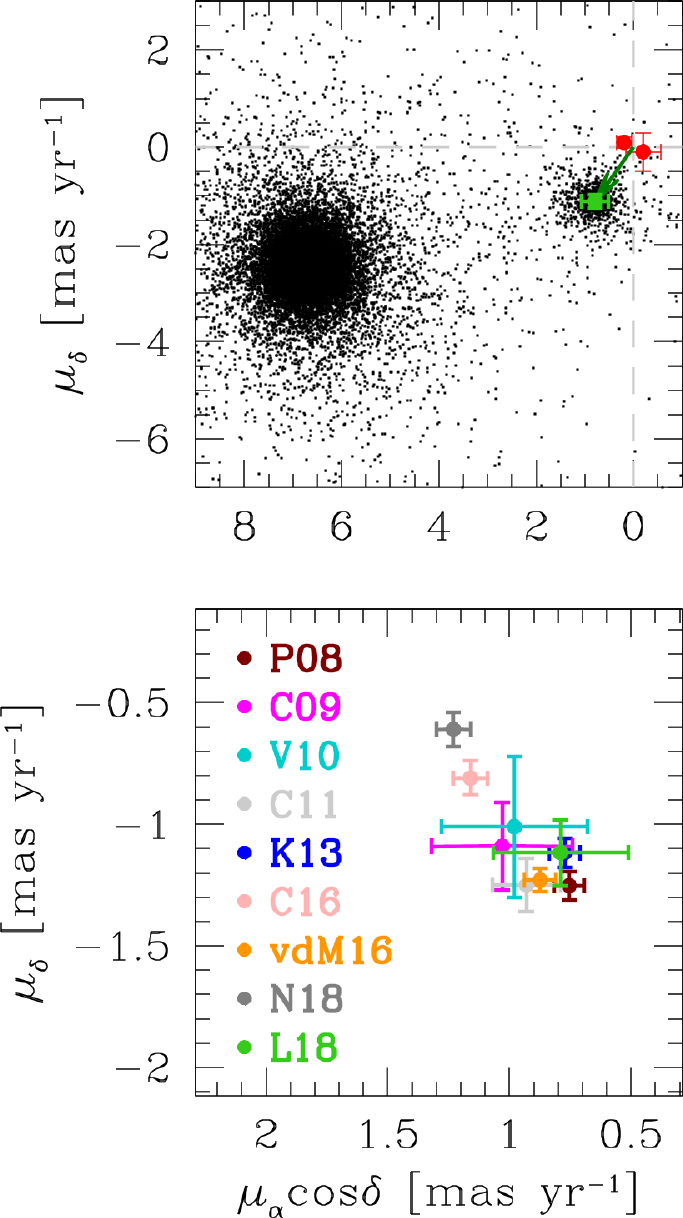}
  \caption{(Top): VPD of absolute PMs in equatorial coordinates. Red
    points are the galaxies used to compute the absolute PM as in
    Fig.~\ref{fig:gal11}. The SMC absolute motion relative to the
    galaxies and the absolute PM value of the SMC are depicted in dark
    and light green, respectively.  (Bottom): Comparison between our
    estimate of the absolute PM of the SMC (L18, green) and literature
    values: \citet[P08, brown]{2008AJ....135.1024P}, \citet[C09,
      magenta]{2009AJ....137.4339C}, \citet[V10,
      turquoise]{2010AJ....140.1934V}, \citet[C11, light
      gray]{2011AJ....141..136C}, \citet[K13,
      blue]{2013ApJ...764..161K}, \citet[C16,
      pink]{2016A&A...586A..77C}, \citet[vdM16,
      orange]{2016ApJ...832L..23V}, \citet[N18, dark
      gray]{2018arXiv180107738N}.}
  \label{fig:gal12}
\end{figure}

The absolute PM of the SMC was computed by using the absolute PM value
of NGC~362 in Sect.~\ref{apm}. We calculated the 4$\sigma$-clipped
median value of the relative PM of SMC stars along each direction. We
estimate:
\begin{equation}
  \begin{array}{ll}
    (\mu_\alpha\cos\delta,\mu_\delta)_{\rm SMC @ NGC~362,\,relative} \\
    = (-5.913\pm0.008,1.291\pm0.007)~{\rm mas~yr^{-1}} \, ,\\
  \end{array}
\end{equation}
thus resulting in:
\begin{equation}
  \begin{array}{ll}
    (\mu_\alpha\cos\delta,\mu_\delta)_{\rm SMC @ NGC~362,\,absolute} \\
    = (0.790\pm0.279,-1.116\pm0.135)~{\rm mas~yr^{-1}} \, .\\
  \end{array}
\end{equation}
Error bars in the latter PM values are the sums in quadrature of the
errors of the absolute PM of NGC~362 and of the relative bulk PM of
SMC stars in our field. The expected contribution of the parallax (see
discussion in Sect.~\ref{apm}) for the stars in our FoV and with the
temporal coverage of our observations is of 10 $\mu$as yr$^{-1}$ along
$\mu_\delta$, while the size of the effect along $\mu_\alpha
\cos\delta$ is negligible. The median PM error of the SMC stars used
in the computation above is $\sim 0.13$ mas yr$^{-1}$, therefore we
chose to not correct for the parallax effects. In the top panel of
Fig.~\ref{fig:gal12}, we show the absolute-PM VPD with the absolute
motion of the SMC highlighted. In the bottom panel of the same Figure,
we present a comparison with the absolute PM of the SMC from the
literature.

The measured PM of SMC stars in the NGC~362 field is not an unbiased
estimate of the PM of the SMC COM. Depending on where one points in
the SMC, different components of the 3D COM velocity vector project
onto the local line of sight, West, and North directions
\citep{2002AJ....124.2639V}. After correcting for viewing perspective
as in \citet{2008ApJ...678..187V}, our new measurement becomes:
\begin{equation}
  \begin{array}{ll}
    (\mu_\alpha\cos\delta,\mu_\delta)_{\rm SMC @ COM,\,absolute} \\
    = (0.741\pm0.279,-1.135\pm0.135)~{\rm mas~yr^{-1}} \, .\\
  \end{array}
\end{equation}
Here the position of the SMC COM is assumed to be at the photometric
centroid of the old stars in the SMC. Note that the viewing
perspective correction is significantly smaller than the random errors
in our measurement.

If one assumes for simplicity that there are no internal motions
within the SMC, then this is an estimate for the PM of the SMC
COM. This result can be compared to existing estimates for the PM of
the SMC COM. \citet{2016ApJ...832L..23V} presented results based on a
combination of \textit{HST} measurements of five fields centered on
background quasars from \citet{2013ApJ...764..161K} and the Gaia DR1
Tycho-Gaia Astrometric Solution (TGAS) catalog measurements of 8
supergiant stars. The results depend on the exact position of the COM,
and the assumed SMC internal kinematics. But with the same assumptions
for these as above, they found that
$(\mu_\alpha\cos\delta,\mu_\delta)_{\rm SMC} = (0.740 \pm 0.072,
-1.202 \pm 0.1070$ mas yr$^{-1}$. Therefore, our new data are
consistent with existing knowledge of the PM of SMC COM, and they do
not significantly improve the existing uncertainties.

Conversely, one could assume that the SMC COM PM is already known from
the literature, and then use our measurement to determine the internal
PM kinematics of the SMC at the position of NGC~362. This yields:
\begin{equation}
  \begin{array}{ll}
    (\mu_\alpha\cos\delta,\mu_\delta)_{\rm SMC,\,int} \\ =
    (0.001\pm0.288,0.067\pm0.172)~{\rm mas~yr^{-1}} \, .\\
  \end{array}
\end{equation}
Given the SMC distance of $\sim 62.8$ kpc, this differs from zero by
20 km s$^{-1}$, but this is well within the uncertainties of 86 km
s$^{-1}$ and 51 km s$^{-1}$, respectively. The measurement
uncertainties are insufficient to probe the internal PM kinematics at
interesting levels. For comparison, the known internal line-of-sight
motions of different stellar populations in the SMC have values
$|\Delta V| \lesssim 30$ km s$^{-1}$ \citep{2014MNRAS.442.1663D}.

%%%%%%%%
\section{The electronic catalog}\label{cat}
%%%%%%%%

The first 10 lines of the astrometric catalog of NGC~362 used in this
paper are presented in Table~\ref{tablea}. The catalog contains both
the original and the corrected PMs. The $X$ and $Y$ positions in our
reference frame (columns 3 and 4) increase toward West and North,
respectively. The pixel scale of our reference frame is set to 40 mas
pixel$^{-1}$ (see Appendix~\ref{red}). The $\chi^2_{\rm X}$ and
$\chi^2_{\rm Y}$ values (columns 9 and 10) are reduced $\chi^2$. The
initial ($N_{\rm f}$) and final ($N_{\rm u}$) number of images
considered in the fit of the PMs are shown in columns (11) and
(12). Column (13) lists the temporal baseline (in yr) adopted to
compute the PMs. The ``ID'' column presents the IDs of the reduction
process.

The first 10 lines of the photometric catalogs of filter F606W and
F814W are shown in Table~\ref{tablev} and Table~\ref{tablei},
respectively. Magnitudes in column (1) are in the Vega-mag system.
The zero-point to subtract to column (1) to obtain the instrumental
magnitudes is 31.7878 and 31.0204 for F606W- and F814W-filter
magnitudes, respectively.

\newpage
\clearpage
\newpage

\begin{sidewaystable}
\caption{First ten lines of the astrometric catalog of NGC~362.}\label{tablea}
\tiny{
\centering
\begin{tabular}{rrrrrrrrrrrrrrrrrr}
\hline\hline
\multicolumn{1}{c}{R.A.} & \multicolumn{1}{c}{Dec.} & \multicolumn{1}{c}{X} & \multicolumn{1}{c}{Y} & \multicolumn{1}{c}{$\!\!\!\mu_\alpha^{\rm r} \cos\delta$} & \multicolumn{1}{c}{$\sigma_{\mu_\alpha^{\rm r} \cos\delta}$}& \multicolumn{1}{c}{$\!\!\!\mu_\delta^{\rm r}$} & \multicolumn{1}{c}{$\sigma_{\mu_\delta^{\rm r}}$}& \multicolumn{1}{c}{$\chi^2_{\rm X}$} & \multicolumn{1}{c}{$\chi^2_{\rm Y}$} & \multicolumn{1}{c}{$N_{\rm f}$}& \multicolumn{1}{c}{$N_{\rm u}$}& \multicolumn{1}{c}{$\!\!\!\Delta$time}& \multicolumn{1}{c}{$\!\!\!\mu^{\rm c}_\alpha \cos\delta$} & \multicolumn{1}{c}{$\sigma_{\mu^{\rm c}_\alpha \cos\delta}$} & \multicolumn{1}{c}{$\!\!\!\mu^{\rm c}_\delta$} & \multicolumn{1}{c}{$\sigma_{\mu^{\rm c}_\delta}$} & \multicolumn{1}{c}{$\!\!\!$ID}\\
\multicolumn{1}{c}{[deg]}&\multicolumn{1}{c}{[deg]}&\multicolumn{1}{c}{[pixel]}&\multicolumn{1}{c}{[pixel]}&\multicolumn{1}{c}{$\!\!\!\!$[mas yr$^{-1}$]$\!\!\!\!$}&\multicolumn{1}{c}{$\!\!\!\!$[mas yr$^{-1}$]$\!\!\!\!$}&\multicolumn{1}{c}{$\!\!\!\!$[mas yr$^{-1}$]$\!\!\!\!$}&\multicolumn{1}{c}{$\!\!\!\!$[mas yr$^{-1}$]$\!\!\!\!$}&   &   &   &   & \multicolumn{1}{c}{[yr]} &\multicolumn{1}{c}{$\!\!\!\!$[mas yr$^{-1}$]$\!\!\!\!$}&\multicolumn{1}{c}{$\!\!\!\!$[mas yr$^{-1}$]$\!\!\!\!$}&\multicolumn{1}{c}{$\!\!\!\!$[mas yr$^{-1}$]$\!\!\!\!$}&\multicolumn{1}{c}{$\!\!\!\!$[mas yr$^{-1}$]$\!\!\!\!$} \\
\multicolumn{1}{c}{(1)}&\multicolumn{1}{c}{(2)}&\multicolumn{1}{c}{(3)}&\multicolumn{1}{c}{(4)}&\multicolumn{1}{c}{(5)}&\multicolumn{1}{c}{(6)}&\multicolumn{1}{c}{(7)}&\multicolumn{1}{c}{(8)}&\multicolumn{1}{c}{(9)}&\multicolumn{1}{c}{(10)}&\multicolumn{1}{c}{(11)}&\multicolumn{1}{c}{(12)}&\multicolumn{1}{c}{(13)}&\multicolumn{1}{c}{(14)}&\multicolumn{1}{c}{(15)}&\multicolumn{1}{c}{(16)}&\multicolumn{1}{c}{(17)}&\multicolumn{1}{c}{(18)}\\
\hline
15.79831672    &    -70.88787766  &  5327.1055 &  1480.9898   &    -0.26052   &    0.08024  &    -0.54896   &    0.08344  &     0.6390  &    0.6665 &  38 & 35  &      2.57836    &   -0.09860   &    0.08205   &   -0.45612   &    0.08616   &   54 \\
15.80203538    &    -70.88771296  &  5217.5273 &  1495.8303   &    -0.25716   &    0.06636  &    -0.26296   &    0.08136  &     0.5899  &    0.7505 &  40 & 40  &      2.57854    &   -0.09900   &    0.06858   &   -0.19420   &    0.08519   &   56 \\
15.80326673    &    -70.88748662  &  5181.2441 &  1516.2059   &    -0.15612   &    0.07244  &    -0.11824   &    0.07780  &     0.5392  &    0.7201 &  39 & 39  &      2.57854    &    0.00204   &    0.07448   &   -0.06984   &    0.08128   &   58 \\
15.79475653    &    -70.88712956  &  5432.0317 &  1548.2969   &    -0.20572   &    0.05192  &    -0.32872   &    0.06528  &     0.5860  &    0.6151 &  38 & 37  &      2.57854    &   -0.04380   &    0.05463   &   -0.23588   &    0.06865   &   62 \\
15.80591215    &    -70.88664623  &  5103.2925 &  1591.8478   &    -0.18976   &    0.05104  &    -0.02756   &    0.05064  &     0.5734  &    0.5808 &  41 & 39  &      2.57854    &   -0.03160   &    0.05395   &    0.00984   &    0.05541   &   73 \\
15.80430153    &    -70.88647132  &  5150.7578 &  1607.5864   &    -0.06992   &    0.06220  &    -0.29952   &    0.07420  &     0.4466  &    0.5938 &  41 & 41  &      2.57854    &    0.08620   &    0.06470   &   -0.25112   &    0.07769   &   75 \\
15.79880840    &    -70.88637743  &  5312.6401 &  1616.0147   &    -0.16712   &    0.05704  &    -0.18592   &    0.04836  &     0.4953  &    0.3460 &  40 & 39  &      2.57854    &   -0.01728   &    0.05927   &   -0.10728   &    0.05212   &   77 \\
15.80746957    &    -70.88586553  &  5057.3984 &  1662.1138   &    -0.19472   &    0.05832  &    -0.10248   &    0.04460  &     0.7451  &    0.4012 &  46 & 46  &      2.57854    &   -0.02760   &    0.06049   &   -0.06624   &    0.04772   &   84 \\
15.80735085    &    -70.88577903  &  5060.8975 &  1669.8987   &    -0.29044   &    0.05360  &     0.00016   &    0.07296  &     0.4045  &    0.6854 &  46 & 45  &      2.57854    &   -0.12332   &    0.05596   &    0.03640   &    0.07491   &   87 \\
15.78989053    &    -70.88497934  &  5575.4883 &  1741.7771   &    -0.14652   &    0.06244  &    -0.09924   &    0.12620  &     0.4498  &    1.0235 &  38 & 37  &      2.57854    &   -0.07644   &    0.06443   &   -0.07300   &    0.12720   &   98 \\
\hline
\end{tabular}
\flushleft \tablecomments{Columns (5), (6), (7), and (8) refer to the
  original, uncorrected PMs. The PMs and corresponding error after
  applying the corrections described in Appendix~\ref{corr} are listed
  in columns (14), (15), (16), and (17).}}
\end{sidewaystable}

\begin{table}
\caption{First ten lines of the photometric catalog of NGC~362 for filter F606W.}\label{tablev}
\tiny{
\centering
\begin{tabular}{rrrrrrrr}
\hline\hline \multicolumn{1}{c}{$m$} & \multicolumn{1}{c}{RMS} &
\multicolumn{1}{c}{\textsf{QFIT}} & \multicolumn{1}{c}{$o$} &
\multicolumn{1}{c}{${\tt RADXS}$} & \multicolumn{1}{c}{$N_{\rm f}$} &
\multicolumn{1}{c}{$N_{\rm u}$} & \multicolumn{1}{c}{RMS sky in
  counts}\\
\hline
     19.5180   &    0.1265  &     0.999   &  0.00051  &     0.0009  &  1  &  1  &   235.91 \\
     19.4744   &    0.0030  &     1.000   &  0.00068  &    -0.0034  &  2  &  2  &   268.41 \\
     19.4463   &    0.0025  &     1.000   &  0.00002  &     0.0013  &  2  &  2  &   282.98 \\
     18.8895   &    0.0727  &     1.000   &  0.00032  &     0.0007  &  1  &  1  &   352.13 \\
     18.4073   &    0.0151  &     1.000   &  0.00000  &     0.0019  &  3  &  3  &   636.99 \\
     19.6737   &    0.0069  &     1.000   &  0.00001  &    -0.0021  &  3  &  3  &   199.27 \\
     19.1991   &    0.0087  &     0.999   &  0.00025  &     0.0043  &  2  &  2  &   292.53 \\
     19.2239   &    0.0158  &     1.000   &  0.00734  &     0.0013  &  4  &  4  &   294.71 \\
     19.5399   &    0.0192  &     1.000   &  0.01133  &    -0.0010  &  4  &  4  &   260.50 \\
     19.7746   &    0.1579  &     0.999   &  0.00000  &     0.0029  &  1  &  1  &   178.40 \\
\hline
\end{tabular}
\tablecomments{The parameter $r_{\rm N}$ described in
  Appendix~\ref{red} can be computed as the ratio between the number
  of individual exposures used to measure a stellar position and flux
  ($N_{\rm u}$) and the total number of exposures that star was
  actually found ($N_{\rm f}$).}}
\end{table}

\begin{table}
\caption{First ten lines of the photometric catalog of NGC~362 for filter F814W.}\label{tablei}
\tiny{
\centering
\begin{tabular}{rrrrrrrr}
\hline\hline
\multicolumn{1}{c}{$m$} & \multicolumn{1}{c}{RMS} & \multicolumn{1}{c}{\textsf{QFIT}} & \multicolumn{1}{c}{$o$} & \multicolumn{1}{c}{${\tt RADXS}$} & \multicolumn{1}{c}{$N_{\rm f}$} & \multicolumn{1}{c}{$N_{\rm u}$} & \multicolumn{1}{c}{RMS sky in counts}\\
\hline
     19.0207   &    0.1598   &    1.000   &  0.00045   &    0.0026  &  1 &   1   &  189.99 \\
     18.9887   &    0.0072   &    1.000   &  0.00102   &   -0.0003  &  2 &   2   &  223.30 \\
     18.9464   &    0.0040   &    1.000   &  0.00004   &    0.0014  &  2 &   2   &  217.67 \\
     18.4258   &    0.0952   &    1.000   &  0.00060   &   -0.0024  &  1 &   1   &  362.10 \\
     17.9270   &    0.0074   &    1.000   &  0.00001   &    0.0006  &  3 &   3   &  564.56 \\
     19.1586   &    0.0092   &    1.000   &  0.00005   &   -0.0005  &  3 &   3   &  177.26 \\
     18.7183   &    0.0075   &    1.000   &  0.00043   &    0.0015  &  2 &   2   &  252.12 \\
     18.7398   &    0.0175   &    0.999   &  0.00940   &    0.0007  &  4 &   4   &  243.20 \\
     19.0367   &    0.0126   &    1.000   &  0.01371   &   -0.0010  &  4 &   4   &  210.64 \\
     19.2601   &    0.1958   &    1.000   &  0.00000   &    0.0027  &  1 &   1   &  163.91 \\
\hline
\end{tabular}}
\end{table}

\end{document}